\documentclass[lettersize,journal]{IEEEtran}
\usepackage{amsmath,amsfonts}
\usepackage{algorithmic}
\usepackage{algorithm}
\usepackage{array}
\usepackage{textcomp}
\usepackage{stfloats}
\usepackage{url}
\usepackage{cuted}
\usepackage{verbatim}
\usepackage{hyperref}
\usepackage{multirow}
\usepackage{pgf,tikz}
\usepackage{enumitem}
\usepackage{float}
\usepackage{comment}
\usepackage{amsfonts}
\usepackage{algorithm}
\usepackage[font=normalsize]{caption}
\usepackage{graphicx}
\usepackage{amsmath,amsthm,amssymb,mathtools,graphicx,cite,calc,color,dsfont,psfrag} 
\usepackage{array}
\usepackage{bm}
\usepackage{enumitem}
\usepackage{subfigure}
\usepackage{hyperref}
\usepackage{xcolor}
\usepackage{cite}
\usepackage{array}
\usepackage{tabularray}
\usepackage{siunitx}
\usepackage{diagbox}
\usepackage{footnote}
\usepackage{tablefootnote}
\usepackage[mathscr]{euscript}
\makeatletter
\newcommand{\thickhline}{%
    \noalign {\ifnum 0=`}\fi \hrule height 1pt
    \futurelet \reserved@a \@xhline
}
\newcolumntype{"}{@{\hskip\tabcolsep\vrule width 1pt\hskip\tabcolsep}}
\makeatother

\begin{document}

\title{HybridChain: Fast, Accurate, and Secure Transaction Processing with Distributed Learning}

\author{Amirhossein Taherpour\,, and Xiaodong Wang\,,~\IEEEmembership{Fellow,~IEEE}
        % <-this % stops a space
\thanks{Electrical Engineering Department, Columbia University\\E-mails: at3532@columbia.edu, xw2008@columbia.edu}% <-this % stops a space
}
% The paper headers
\markboth{}%
{}

\IEEEpubid{}
% Remember, if you use this you must call \IEEEpubidadjcol in the second
% column for its text to clear the IEEEpubid mark.

\maketitle
In order to fully unlock the transformative power of distributed ledgers and blockchains, it is crucial to develop innovative consensus algorithms that can overcome the obstacles of security, scalability, and interoperability, which currently hinder their widespread adoption. This paper introduces HybridChain that combines the advantages of sharded blockchain and DAG distributed ledger, and a consensus algorithm that leverages decentralized learning. Our approach involves validators exchanging perceptions as votes to assess potential conflicts between transactions and the witness set, representing input transactions in the UTXO model. These perceptions collectively contribute to an intermediate belief regarding the validity of transactions. By integrating their beliefs with those of other validators, localized decisions are made to determine validity. Ultimately, a final consensus is achieved through a majority vote, ensuring precise and efficient validation of transactions. Our proposed approach is compared to the existing DAG-based scheme IOTA and the sharded blockchain Omniledger through extensive simulations. The results show that IOTA has high throughput and low latency but sacrifices accuracy and is vulnerable to orphanage attacks especially with low transaction rates. Omniledger achieves stable accuracy by increasing shards but has increased latency. In contrast, the proposed HybridChain exhibits fast, accurate, and secure transaction processing, and excellent scalability. 

\section{Introduction} \label{sec:intro}
Distributed ledgers, commonly known as blockchains, have emerged as disruptive technologies capable of revolutionizing various industries. Such decentralized and transparent systems enable secure and immutable recording of transactions without the need for intermediaries such as government agencies. Built on a peer-to-peer network of computers that collectively validate and record transactions, the blockchains employ cryptographic techniques to ensure the integrity and authenticity of the data. Originally developed to create and exchange digital currencies such as Bitcoin~\cite{QQ1}, blockchains technology now find applications in various domains, making transformative impacts~\cite{QQ2}.\par
However, the potential benefits of the blockchain technology are accompanied by challenges, particularly in scalability~\cite{QQ3}. Blockchain systems have limited capacity and can process only a limited number of transactions per time unit. This issue is challenging to address due to the blockchain trilemma~\cite{QQQ1}, which refers to the difficulty of achieving high levels of scalability, security, and decentralization simultaneously. Any improvement in one aspect would come at the expense of one or both of the others, and finding the right balance among the three aspects is critical for the success of a blockchain system.\par
Solutions to addressing these issues can be broadly classified as either off-chain or on-chain~\cite{QQQ2}. Off-chain solutions improve the scalability by reducing the computational burden on the main blockchain. They include state channels~\cite{QQQ3}, payment channels~\cite{QQQ4}, sidechains~\cite{QQQ5}, and Plasma~\cite{QQQ6}, all of which handle transactions off-chain to reduce congestion on the main chain and thus increase the scalability of the blockchain. On the other hand, on-chain solutions involve changes to the underlying blockchain protocol, such as changing the block data structure, and employing a new consensus mechanism. In particular, AdaptChain~\cite{QQ8}, Jidar~\cite{QQ9}, and CUB~\cite{QQ10} compress the size of transaction representation in each block, saving the bandwidth and reducing the propagation delay, thereby increasing the throughput. On the other hand, Algorand~\cite{QQ11} and~\cite{QQ13} each propose a new consensus algorithm based on Pure Proof of Stake (PPoS) and Votes-as-a-Proof (VaaP), respectively, to increase the throughput.\par
Moreover, as an on-chain solution, sharding which is adopted in Meepo~\cite{QQ5}, Benzene~\cite{QQ6}, Omniledger~\cite{QQQ2000}, and~\cite{QQ7}, is the most significant approach among all on-chain and off-chain approaches. Firstly, sharding partitions the network into smaller shards, and each shard operates independently, processing its own subset of transactions and smart contracts. This division of labor reduces the computational burden on individual nodes and increases the overall throughput of the network. Off-chain solutions like state channels and payment channels can help reduce congestion on the main chain but have limited scalability potential. The reason for this limitation is that these solutions depend on direct communication and coordination among the participants, which imposes a cap on the number of nodes that can participate.\par
Secondly, sharding maintains the security and decentralization benefits inherent in blockchains. Each shard has its own set of validators, ensuring that consensus is reached on the subset of transactions it processes. By distributing the network's computational load across multiple shards, sharding enhances the overall network's resilience against attacks and reduces the risk of a single point of failure. In comparison, off-chain solutions like sidechains and Plasma often introduce additional trust assumptions and require trust in a set of federated entities or a single trusted party, compromising the decentralized nature of the blockchain.\par
Furthermore, sharding facilitates efficient data storage and retrieval. With the partitioning of the blockchain into smaller shards, each shard only needs to store a fraction of the entire blockchain's data, reducing storage requirements for individual nodes. Additionally, sharding enables parallel processing and faster block propagation within shards, resulting in reduced latency and faster transaction confirmation. On-chain solutions like AdaptChain, Jidar, and CUB offer optimization techniques to improve throughput, but they do not provide the same level of scalability as sharding.\par
On the other hand, a new avenue gaining significant attention is the Directed Acyclic Graph (DAG) approach. Unlike existing off-chain and on-chain solutions designed for blockchains, DAG is a unique data structure that enables non-linear data recording in a distributed ledger. Transactions in a DAG are interlinked based on their dependencies, forming a directed graph with no cycles. The structure permits parallel processing of transactions, providing a higher throughput than conventional blockchain systems. DAG-based distributed ledgers could address scalability and transaction processing challenges for two reasons. Firstly, as the number of users increases in the network, the transaction processing capacity of a DAG can increase, whereas conventional blockchain systems become less efficient. Secondly, DAGs can offer quicker transaction confirmation compared to blockchains. Transactions in a DAG can be confirmed as soon as they are added to the network without waiting to be processed in a block, as in a blockchain. DAG schemes include Nano~\cite{QQ15}, Avalanche~\cite{QQ16}, Graphchain~\cite{QQ17}, Phantom~\cite{QQ18} and IOTA~\cite{QQQ2001}.\par
However, despite the advantages of the sharded blockchains and DAG distributed ledgers, they have a number of drawbacks~\cite{QQ19, QQ20} that will be elaborated in Section II-B. Motivated by these drawbacks, in this paper, we propose a novel consensus mechanism that provides quick transaction processing like DAG but with improved accuracy and security, achieved by introducing validators as in blockchain. This results in a scalable, efficient system capable of processing high volumes of transactions with high accuracy.\par
The main contributions of this paper are as follows:
\begin{enumerate}
    \item We propose a new transaction validation scheme that prioritizes the transactions more likely to be valid for faster and more efficient processing, thus achieving higher throughput.
    \item We propose a lightweight consensus algorithm that utilizes decentralized learning. Validators exchange perceptions to assess transaction conflicts with the witness set, representing input transactions in the UTXO model. These perceptions form an intermediate belief about validity. Validators combine beliefs with each others, leading to local decisions on validity. Final consensus emerges through a majority vote, ensuring accurate and efficient transaction validation.
    \item We discuss in-depth the notions of security, decentralization, and throughput and compare our proposed scheme with IOTA and Omniledger in terms of these metrics.
\end{enumerate}
The remainder of the paper is organized as follows. Section II provides a brief overview of the conventional blockchain, sharded blockchain, and the DAG distributed ledger. Section III presents our proposed HybridChain scheme. Section IV compares our scheme with IOTA and Omniledger in terms of security. Section V, compares our scheme with IOTA and Omniledger in terms of scalability of throughput and latency through simulations. Finally, Section VI concludes the paper.
\section{Backgrounds}
This section first gives an overview of the conventional blockchain, sharded blockchain, and DAG distributed ledger. Then, we discuss the drawbacks of these existing schemes.\par
\subsection{Existing Approaches}
\subsubsection{Conventional Blockchain}
Distributed ledger technologies utilize blockchains to store ordered records called blocks, facilitated by a network of untrusted entities known as validators. These blocks are linked through references to previous block headers, creating an immutable chain. A typical block consists of two main sections: transactions and header. The header contains the hash of the previous block header and the hash of the Merkle root, which represents the digest of all transactions in the block. Validators propose new blocks using a consensus algorithm, selecting transactions and forming the block header. When creating the next block, validators verify the validity of the proposed block, ensuring it adheres to the underlying protocols. Once confirmed, the validators accept the proposed block as the next block, appending it to the existing chain.\par
The consensus mechanism used in a blockchain determines how blocks are proposed. There are primarily two types of consensus mechanisms: Nakamoto Consensus and Classical Consensus~\cite{QQ007}. Nakamoto Consensus includes Proof of Work (PoW) and Proof of X (PoX), where miners solve mathematical problems or utilize a scarce resource to propose blocks. Classical Consensus, on the other hand, relies on voting and allows for faster confirmation of transactions; Practical Byzantine Fault Tolerance (PBFT) is an example where nodes communicate in quorums to agree on the system's state. Note that Nakamoto Consensus can result in wasted efforts, while classical consensus may suffer from congestion and low throughput. To address these issues, solutions like sharded blockchain and DAG distributed ledgers have been proposed. \par
Next we provide an overview of Omniledger~\cite{QQQ2000}, which is a sharded blockchain, and IOTA ~\cite{QQQ2001}, which is a distributed ledger based on DAG.\par

\subsubsection{Sharded blockchain — Omniledger} In a conventional blockchain, all validators validate and store every transaction. On the other hand, sharding partitions the blockchain into smaller subsets called shards, where each shard is responsible for processing and storing a subset of transactions. By distributing the workload across validators, sharding enables parallel processing and increases the throughput. Unlike non-sharded blockchains, where every validator processes all transactions, sharding allows validators within a shard to focus solely on the transactions in their shard, resulting in faster confirmation times. Additionally, sharding facilitates scalability by adding more shards as the network expands. However, sharding also introduces new challenges such as shard coordination, secure cross-shard communication, and specialized consensus protocols.\par
As a sharded blockchain, Omniledger operates in epochs and each epoch consists of four main stages. Initially, validators are assigned to shards, and leaders are selected using the \emph{RandHound} distributed randomness beacon. In the second stage, these leaders propose new blocks containing transactions limited to their respective shards. Validation of the proposed block within each shard is conducted by validators, employing the \emph{ByzCoinX} consensus algorithm, followed by voting on its validity. Once a two-thirds majority of validators agree, the block is added to the blockchain. The third stage addresses cross-shard transactions through the utilization of \emph{Atomix}, ensuring atomicity across multiple shards. In this stage, the initiating shard secures the relevant accounts, creates an Atomix instance, and requests participation from other shards involved in the transaction. Execution of the transaction occurs when all participating shards reach an agreement, and upon success, the Atomix instance is committed. In the final stage, the epoch is concluded, resulting in the update of the blockchain state and preparation for the next epoch.

\subsubsection{DAG — IOTA} In IOTA, the constructed Directed Acyclic Graph (DAG) is referred to as the Tangle, where vertices represent transactions. In the Tangle, a directed edge from vertex $v_i$ to $v_j$ signifies that transaction $v_i$ validates transaction $v_j$ \cite{QQQ2001}.\par
In IOTA, the notion of validators associated with traditional blockchains does not exist. Instead, the ledger relies on a network of nodes. i.e., devices or computers, to submit and process transactions. When nodes submit transactions, they are verified by other nodes within the network. This self-validating mechanism ensures the legitimacy and consistency of transactions. IOTA consists of two types of nodes: light nodes and full nodes. Light nodes are responsible for expanding the Tangle by generating and attaching new transactions to it, while full nodes store all the data in the Tangle. Light nodes rely on full nodes to retrieve the necessary information.\par
Similar to the concept of confirmed blocks that are irrevocable in blockchains, nodes in IOTA must reach consensus on the set of confirmed transactions. To achieve this consensus, for each generated transaction in IOTA, light nodes fetch the required information from full nodes and solve a Proof of Work (PoW) puzzle. They then select two older transactions, called parents, from the tip transaction pool\cite{QQ14}. Here, a tip transaction refers to a transaction that has been broadcast to the network but has not yet been validated by being directly or indirectly referenced by another transaction. This parent selection in the Tangle allows for the existence of two sets of transactions associated with a given transaction: the past cone, which consists of transactions referenced directly or indirectly, and the future cone, which includes transactions that reference the given transaction either directly or indirectly. \par
To finalize the consensus, IOTA employs a trusted node, known as the Coordinator, that periodically issues special transactions called milestones, which act as checkpoints for the Tangle. Transactions that are in the past cone of these milestones are considered confirmed. \par
It is worth noting that since nodes must solve a PoW puzzle to attach the transactions to the Tangle, the consensus mechanism in IOTA is classified as a Nakamoto consensus.\par

\subsection{Drawbacks of Existing Schemes and Motivation of Proposed Approach}
While DAG schemes are capable of decreasing latency by removing validators and processing transactions solely based on their positions in the submission order graph, this results in reduced throughput, especially when there are dishonest nodes that behave strategically. On the other hand, a sharded blockchain, in which validators process transactions in parallel, can improve throughput scalability while keeping latency low in such circumstances.\par
However, the processing of cross-shard transactions poses a significant challenge for sharded blockchains. Specifically, the shards must exchange information about the transaction's validity and the associated account balances to process cross-shard transactions. This necessitates a mechanism for cross-shard communication and consensus, which can be intricate and time-consuming. Moreover, a large volume of cross-shard transactions can potentially create a bottleneck in the system, as the shards must collaborate and communicate effectively to ensure that all transactions are processed accurately and promptly. Omniledger aims to address the challenges associated with cross-shard transactions by implementing Atomix processing. However, as we will demonstrate in Section V, cross-shard transactions in Omniledger can still introduce significant network delays, particularly when the number of shards increases and a larger volume of transactions are submitted as cross-shard.\par
Moreover, the notion of decentralization in a distributed ledger covers both storage and computation. Decentralizing storage involves distributing data across multiple nodes with minimal redundancy so that each node stores only a fraction of the ledger data. Decentralizating computation involves distributing the computing process for transactions among validators in the network.\par
Omniledger exhibits good decentralization of both storage and computation since, as a sharded blockchain, validators process transactions for their respective shards and store the associated data. On the other hand, for IOTA DAG where validators are not present to process transactions, edges of the DAG graph created by nodes in the network validate or reject vertices corresponding to transactions. Decentralization of computations in IOTA can be achieved to some extent by ensuring that the decision to accept or reject a vertex is not made by a single or small group of nodes.\par
Furthermore, IOTA lacks adequate storage decentralization, since light nodes in IOTA heavily depend on full nodes to send or receive required information for selecting transaction parents and updating their perception of the Tangle. Additionally, as explained in Section II-A-2, the decentralization of computations within IOTA is governed by the Coordinator, a centralized entity responsible for determining the set of confirmed transactions. Consequently, it becomes evident that IOTA lacks computational decentralization.\par
Additionally, as explained in Section IV, Omniledger is found to be vulnerable to replay attacks and message withholding attacks. On the other hand, IOTA faces its own set of vulnerabilities, namely orphanage attacks and routing attacks.\par  
Our proposed HybridChain in this paper aims to tackle the aforementioned challenges by adopting a hybrid approach that combines the benefits of both DAG and sharded blockchains. Similar to DAG-based schemes, our scheme has the advantage of low latency and quick transaction processing. Additionally, our scheme features parallel transaction processing by validators, akin to sharded blockchains, resulting in decentralized storage and computations. We also ensure sufficient redundancy in the data storage and employ a consensus mechanism that enables our scheme to withstand attacks that both DAG and sharded blockchains are susceptible to. Table.\ref{sca} summarizes the level of scalability, decentralization, and secutiry of the three schemes.\par
\begin{table}[htb]
\centering
\resizebox{0.48\textwidth}{!}{
    \begin{tblr}{hlines={1pt, black}, vlines={1pt, black},
                 colspec = {Q[c, wd=12em, font=\normalsize  \bfseries] 
                       *{5}{Q[c, wd=8.5em, si={table-format=3}]}},
                 row{1}  = {font=\normalsize \bfseries},
                 rows    = {bg=white!30},
                 rowsep  = {1ex}
                 }
\SetCell[r=2]{c}    \diagbox{Scheme}{Trilemma}  
        & \SetCell[c=2]{c}  {{{Scalability}}}
                &       & \SetCell[c=2]{c}  {{{Decentralization}}}
                                    &           & \SetCell[r=2]{c}  {{{Security}}}   \\
        &{{{\textbf{Throughput}}}} &{{{\textbf{Latency}}}}   &{{{\textbf{Storage}}}} & {{{\textbf{Computation}}}}  &       \\
{IOTA}    & \text{Moderate}   & \text{Good}         & \text{Poor}    &  \text{Moderate}       & \text{Poor}   \\
{Omniledger}  &  \text{Moderate} &   \text{Poor}       & \text{Good}    & \text{Good}        & \text{Poor}   \\
{HybridChain}   & \text{Good}        & \text{Good}             & \text{Good}         & \text{Good}            & \text{Good}  \\
    \end{tblr}}
\caption{\label{sca} Comparison of IOTA, Omniledger, and the proposed HybridChain based on the trilemma of scalability, decentralization, and security.}
\end{table}

\section{HybridChain}
This section describes our proposed HybridChain in detail. We first provide a overview and then describe the transaction attributes that are used by HybridChain. Next, we elaborate on the transaction processing procedure and how the system parameters are updated. Finally, we describe how consensus is reached by validators.\par
\subsection{Overview} 
Suppose that the network consists of $N$ users $U=\{u_{1},\ldots,u_{N}\}$ that submit transactions, and $M$ validators $\mathcal{M}=\{m_{1},\ldots,m_{M}\}$ that process transactions. Note that these two sets can overlap. Corresponding to each validator or user we define a reliability value $\rho_{i}^{m}\in [0,1], i=1, \ldots, M$ or $\rho_{j}^{u}\in [0,1], j=1, \ldots, N$. Each validator can exchange information with all other validators in the network. Each validator is either honest or dishonest: honest validators strictly adhere to all network rules and protocols, while Byzantine dishonest validators act in a random and coordinated fashion. The Byzantine validators possess comprehensive knowledge of the network, including the network structure, the algorithms employed, and the exchanged data. We assume that at most $f\leq\lfloor M/2 \rfloor-1$ validators are Byzantine dishonest.\par
Validators use the UTXO model to store past transactions in the ledger, where each transaction $x_{i}$ is created by some of the past transactions $\{x_{1}, x_{2}, \ldots, x_{i-1}\}$. Denote $W_{i}\subset\{x_{1},\ldots,x_{i-1}\}$ as the witness set of $x_{i}$. That is, $x_{i}$ is valid if it is not in conflict with any of the transactions in $W_{i}$.\par
Similar to the sharded blockchain, we partition time into epochs, but with variable durations. Each epoch comprises several rounds, and the number of rounds within each epoch can vary. In each epoch validators process $\lambda=\lfloor M/(2f+2) \rfloor$ transactions (the choice of this value is explained in Remark 5 in Section III-D). Denote the batch of transactions that is processed during epoch $n$ by $X(n)=\{x_{(n-1)\lambda+1},x_{(n-1)\lambda+2},\ldots,x_{n\lambda}\}$. Then for this epoch, validators are randomly grouped into $\lambda$ communities of size $M/\lambda$ and validators in one community process a single transaction in $x_{\ell}\in X(n)$. Denote the set of validators in the community that is responsible for processing transaction $x_{\ell}\in X(n)$ by $\mathcal{V}_{\ell}$. Then validators in $\mathcal{V}_{\ell}$ process $x_{\ell}$ in maximum $\lvert W_{\ell} \rvert$ rounds, and if $x_{\ell}$ is validated, then it is stored by all validators in $\mathcal{V}_{\ell}$. Since all $\lambda$ transactions in $X(n)$ are processed in parallel across communities, the maximum number of rounds in epoch $n$ is $R_{n}=\max\{\lvert W_{(n-1)\lambda+1} \rvert , \lvert W_{(n-1)\lambda+2} \rvert ,\ldots, \lvert W_{n\lambda} \rvert\}$.\par

In each round $r = 1, \ldots, R_{n},$ validator $m_{k}\in \mathcal{V}_{\ell}$ computes a quantity $p_{k}^{\ell}(r)\in [0,1]$ that represents the current actual belief of validator $m_{k}$ about the validity of transaction $x_{\ell}$. (The calculation of $p_{k}^{\ell}(r)$ is given in Section III-C.) Then the local decision by validator $m_{k}$ at round $r$ about the validity of transaction $x_\ell$, denoted by $v_{k}^{\ell}(r)$, is given by:
\begin{equation}\label{thresholds}
v_{k}^{\ell}(r) = 
\begin{cases}
    1, & \text{if } p_{k}^{\ell}(r) \geq \eta_{1}(\mathbf{a}_{\ell},\mathbf{y}_{k}) = \mu_{1} - \frac{\sigma(\mathbf{a}_{\ell}^{T}\mathbf{y}_{k})}{2}, \\
    0, & \text{if } p_{k}^{\ell}(r) \leq \eta_{2}(\mathbf{a}_{\ell},\mathbf{y}_{k}) = \mu_{2} - \frac{\sigma(\mathbf{a}_{\ell}^{T}\mathbf{y}_{k})}{2},\\
\phi, & \mbox{otherwise},  
\end{cases}
\end{equation}
where $\phi$ denotes that a local decision cannot be made, $\eta_{1}(\mathbf{a}_{\ell},\mathbf{y}_{k})$ and $\eta_{2}(\mathbf{a}_{\ell},\mathbf{y}_{k})$ are the acceptance and rejection thresholds, respectively, with $\mu_{1}$ and $\mu_{2}$ being some fixed values, $\sigma(\cdot)$ being a sigmoid function, $\mathbf{a}_{\ell}$ being an attribute vector for transaction $x_{\ell}$ and $\mathbf{y}_{k}$ being a weight vector associated with validator $m_{k}$.\par
The final collective decision, i.e., the network consensus about the validity of $x_{\ell}$, can be reached at round $r$ if either of the following two conditions is met:
\begin{equation}\label{finaldecision}
V^{\ell} = \begin{cases}
1, & \text{if } \sum_{m_{k}\in \mathcal{V}_{\ell}}\mathbf{1}_{[v_{k}^{\ell}(r)=1]}>\frac{1}{2}\lvert \mathcal{V}_{\ell} \rvert, \\
0, & \text{if } \sum_{m_{k}\in \mathcal{V}_{\ell}} \mathbf{1}_{[v_{k}^{\ell}(r)=0]}>\frac{1}{2}\lvert \mathcal{V}_{\ell} \rvert,
\end{cases}
\end{equation}
where $V^{\ell}=1$ means that transaction $x_{\ell}$ is confirmed by the network and $V^{\ell}=0$ otherwise, and $\mathbf{1}_{[\cdot]}$ is the indicator function.\par

\subsection{Attribute Vector}
Here we discuss the idea behind using the attribute vector $\mathbf{a}_{\ell}$ for transaction $x_{\ell}$. Unlike existing distributed ledger schemes that spend the same effort to process all submitted transactions, our scheme adjusts the time and computational resources. The idea is that all submitted transactions do not need the same amount of processing, and some are more likely to be valid than others. Therefore, transactions that are more likely to be valid can be treated less strictly to save time and other resources. In this regard, we use the attribute vector $\mathbf{a}_{\ell}$ for transaction $x_{\ell}$ to find its threshold of acceptance or rejection. The elements of $\mathbf{a}_{\ell}$, along with their descriptions, are listed in Table \ref{att}.

\begin{table}[htb]
\centering
%\begin{adjustbox}{angle=90}
\resizebox{0.48\textwidth}{!}{
\begin{tabular}{|c"c|c|c|}
 \hline
\textbf{Notation} &  \textbf{Description}   \\ \thickhline
$a_{\ell}[1]$ & The inverse of the value of the asset that is transferred from sender(s) to receiver(s) \\[5pt]\hline 
$a_{\ell}[2]$ & The amount of the fee that a transaction submitter pays to validators for processing\\[5pt] \hline 
$a_{\ell}[3]$ & The elapsed time from the previous transaction submitted by the same user in number of rounds \\[5pt] \hline
$a_{\ell}[4]$ & The inverse of the witness set size\\ [5pt]\hline 
$a_{\ell}[5]$        & The weighted average of reliability of the user(s) that initiate the transaction \\ [5pt]\hline   
\end{tabular}}
%\end{adjustbox}
\caption{\label{att} Elements of the attribute vector $\mathbf{a}_{\ell}$.}
\end{table}  
Here we explain the intuition behind the relation between each element of $\mathbf{a}_{\ell}$ and the validity of transaction $x_{\ell}$. For each element of $\mathbf{a}_{\ell}$, a higher value indicates a higher likelihood of $x_{\ell}$ being valid. Consequently, the acceptance and rejection thresholds for transaction $x_{\ell}$ in (\ref{thresholds}) will be lower.\par
$a_{\ell}[1]$ is the inverse of the value of the transaction $x_{\ell}$. A transaction triggering a larger amount of asset transfer between two users should be processed more carefully. However, while comparing two transactions, especially the ones that have lesser values, the transaction with higher fees can decrease the likelihood of invalid transaction. Hence $a_{\ell}[2]$ is the fee that must be paid to the validators for processing transaction $x_{\ell}$ regardless whether $x_{\ell}$ will be confirmed or not. $a_{\ell}[3]$ is the elapsed time from the previous transaction submitted by the same user in number of rounds. This means if one user submits many transactions in a short time, it can be a sign of abnormality. $a_{\ell}[4]$ is the inverse of the size of the witness set $W_{\ell}$ for transaction $x_{\ell}$. Since a transaction with a bigger witness size $\lvert W_{\ell}\rvert$ is triggering a bigger number of unspent transactions $x_{j}\in W_{\ell}$, then such transaction should be handled more prudently. In general, transaction $x_{\ell}$ can be initiated by multiple users from the set $U$. This means that each transaction in the witness set $W_{\ell}$ can have a different owner, which can be considered analogous to a multi-signature transaction in the UTXO model. Specifically, for each $a_{\ell}[5]$, represents the weighted average of the reliability of the sender(s) transaction $x_{j}\in W_{\ell}$. Let $\mathcal{U}(x_{j})$ denote the index of the user who owns $x_{j}$, and $\mathcal{A}(x_{j})$ represent the value of $x_{j}$. Then $a_{\ell}[5]=\sum\limits_{x_{j}\in W_{\ell}}\frac{\mathcal{A}(x_{j})}{\sum\limits_{x_{j}\in W_{\ell}}\mathcal{A}(x_{j})}\rho_{\mathcal{U}(x_{j})}^{u}$.\par

%####################################################################################################################################################################################################################################################################
\subsection{Transaction Validation Procedure}
Recall that in each epoch $n$, each transaction $x_{\ell} \in X(n)$, as well as information on its witness set $W_{\ell}$ is available to all validators. Therefore, for any $x_{j}\in W_{\ell}$, if validator $m_{k} \in \mathcal{V}_{j}$, then it can determine whether or not $x_{\ell}$ and $x_{j}$ are in conflict, and we define the perception:
\begin{equation}
q_{k}(x_{\ell},x_{j}) = \begin{cases}
0, & \text{if $x_{\ell}$ and $x_{j}$ are in conflict} \\
1, & \text{if $x_{\ell}$ and $x_{j}$ are not in conflict};
\end{cases}
\end{equation}
otherwise validator $m_{k}$ cannot discern. Moreover, denote by $W_{\ell}^{k}$ a random permutation of the elements in $W_{\ell}$ generated by validator $m_{k}$. That is, for two validators $m_{k}$ and $m_{k^{\prime}}$, $W_{\ell}^{k}$ and $W_{\ell}^{k^{\prime}}$ have the same elements but with different orderings. We denote the transaction corresponding to the $r$-th element in $W_{\ell}^{k}$ by $W_{\ell}^{k}(r)$.\par
\subsubsection{Processing procedure in each round $r$}
After each round $r$ in epoch $n$, one or more transactions may be eliminated from the set of the transactions $X(n)$ since their processing is finished by their corresponding validators, i.e., $v_{k}^{\ell}(r)=0$ or $1$ in (\ref{thresholds}). We denote $X_{r}(n)$ as the set of transactions still being processed after round $r$ of epoch $n$ with $X(n)=X_{1}(n)$. Then for each round $r$ in epoch $n$, the processing of all transactions in $X_{r}(n)$ consists of the following four stages:\par

\textbf{Stage 1: Validators in all communities exchange perceptions:}\par
\emph{Step 1: Sending perceptions to other validators:} For each transaction $x_{\ell}\in X_{r}(n)$, and for all transactions $x_{\ell^{\prime}}\in X_{1}(n)$ each validator $m_{\kappa}\in \mathcal{V}_{\ell}$ checks if $r\leq \lvert W_{\ell^{\prime}} \rvert$ and $m_{\kappa}\in\mathcal{V}_{W_{\ell^{\prime}}^{\kappa}(r)}$. If so it sends $q_{k}(x_{\ell^{\prime}},W_{\ell^{\prime}}^{\kappa}(r))$ to all validators in community $\mathcal{V}_{\ell^{\prime}}$. Otherwise, $m_{\kappa}$ does not send any data regarding transaction $x_{\ell^{\prime}}$.\par
 \emph{Step 2: Receiving perceptions from other validators:} For each transaction $x_{\ell}\in X_{r}(n)$, each validator $m_{k}\in \mathcal{V}_{\ell}$ receives $q_{k}(x_{\ell},W_{\ell}^{\kappa}(r))$ that are sent in the Step 1 from all other validators $m_{\kappa}$. We denote the set of all such perceptions by $Q_{k}^{\ell}(r)$.\par
\textbf{Stage 2: Update actual belief and making local decision by each validator:}\par
\emph{Step 1: Updating actual belief:} For each transaction $x_{\ell}\in X_{r}(n)$, each validator $m_{k}\in \mathcal{V}_{\ell}$ updates $p_{k}^{\ell}(r-1)$ using $Q_{k}^{\ell}(r)$ received at Stage 1 to obtain $p_{k}^{\ell}(r)$. The details can be found in Section III-C.\par 
\emph{Step 2: Making local decisions:} Validator $m_{k}$ calculates $v_{k}^{\ell}(r)$ in (\ref{thresholds}).\par
\textbf{Stage 3: Reach collective decision within each community $\mathcal{V}_{\ell}$:}\par
\emph{Step 1: Sending local decision to other validators:}
Each validator $m_{k}\in \mathcal{V}_{\ell}$ that has made a decision in Stage 2 sends $v_{k}^{\ell}(r)\in \{0,1\}$ to all other validators in $\mathcal{V}_{\ell}$.\par  
\emph{Step 2: Receiving local decisions from other validators:} Each validator $m_{k}\in \mathcal{V}_{\ell}$ receives $v_{k^\prime}^{\ell}(r)$ from other validators $m_{k^\prime}\in \mathcal{V}_{\ell}$.\par
\emph{Step 3 (a): Reaching collective decision:} Each validator $m_{k}\in \mathcal{V}_{\ell}$ forms $\mathcal{M}^{\ell}(r)=\{m_{k^\prime}\mid m_{k^\prime}\in \mathcal{V}_{\ell},\,\, v_{k^\prime}^{\ell}(r^{\prime})\in \{0,1\} \,\,\text{for}\,\, 1\leq r^{\prime}\leq r\}$, i.e., the set of  validators in community $\mathcal{V}_{\ell}$ that have made local decisions about the validity of $x_{\ell}$ up to round $r$. Then validator $m_{k}$ checks if there is at least $\lfloor \frac{\lvert \mathcal{V}_{\ell} \rvert}{2}\rfloor+1$ identical local decisions in $\mathcal{M}^{\ell}(r)$. If so, the final decision $V^\ell$ regarding the validity of transaction $x_\ell$ will be based on this agreement.\par 
\emph{Step 3 (b): Forcing validators to make a decision:} If there are less than $\lfloor \frac{\lvert \mathcal{V}_{\ell} \rvert}{2}\rfloor+1$ identical decisions in $\mathcal{M}^{\ell}(r)$ and if $r=\lvert W_{\ell} \rvert$, then each validator $m_{k}\in \mathcal{V}_{\ell}\backslash \mathcal{M}^{\ell}(r)$ makes a decision based on the following rule:
\begin{equation}
\eta_{1}(\mathbf{a}_{\ell},\mathbf{y}_{k})- p_{k}^{\ell}(r=\lvert W_{\ell} \rvert)\underset{v_{k}^{\ell}=1}{\overset{v_{k}^{\ell}=0}{\gtrless}} p_{k}^{\ell}(r=\lvert W_{\ell} \rvert)-\eta_{2}(\mathbf{a}_{\ell},\mathbf{y}_{k}).
\end{equation} 
This forces all validators in community $\mathcal{V}_{\ell}$ to make local decisions by round $r=\lvert W_{\ell} \rvert$ and the final decision can be obtained using (\ref{finaldecision}).\par   
\textbf{Stage 4: Broadcasting final decisions:} For each community $\mathcal{V}_{\ell}$, if a final decision $V^{\ell}$ on transaction $x_{\ell}$ is reached in the current round $r$, then $V^{\ell}$ is broadcast to all validators in $X(n)$. Each validator then obtains $X_{r+1}(n)$ by removing all transactions whose final decisions have been reached in round $r$.\par    
\emph{Remark 1:} Although the final collective decision for transaction $x_{\ell}$ must be made by round $\lvert W_{\ell} \rvert$, it can be reached sooner if sufficient number of validators in community $\mathcal{V}_{\ell}$ reach local decisions before round $\lvert W_{\ell} \rvert$. Therefore, all processings in epoch $n$ can be completed before round $R_{n}$ if every community reaches a final decision earlier than round $\lvert W_{\ell} \rvert$.\par     
\emph{Remark 2:} Note that if the final decision $V^{\ell}$ for transaction $x_\ell$ is reached in round $r$, then in the remaining rounds $r^\prime>r$ of the epoch validators in $\mathcal{V}_{\ell}$ participate only in Step 1 of Stage 1 and Stage 4.\par
\emph{Remark 3:} The HybridChain can be viewed as a hybrid of the sharded blockchain and the DAG. In particular, it is evident that validators process transactions concurrently in $\lambda$ communities, resembling sharded blockchains. Furthermore, akin to DAG distributed ledgers, HybridChain utilizes thresholds as defined in (\ref{thresholds}) to determine the acceptance or rejection of transactions. Additionally, similar to DAG where the relationships among transactions dictate the set of valid transactions, HybridChain assesses the validity of a transaction based on its confliction to the transactions within its witness set.\par
\emph{Remark 4:} The decentralized nature of HybridChain is apparent in its implementation. Specifically, each transaction $x_{\ell}$, is stored by $2f+2$ validators in the set $\mathcal{V}_{\ell}$, which ensures a decentralized ledger. Furthermore, the final decision $V^{\ell}$ is not solely determined by local decisions of validators in $\mathcal{V}_{\ell}$, but also influenced by perceptions received from all other validators in different communities $\mathcal{V}_{\ell^\prime}$ during Stage 1. This decentralized approach to processing $x_{\ell}$ ensures that computations related to its validation are decentralized across the network.

\subsubsection{Updating reliability and weight in each epoch $n$}
At the end of each epoch $n$, the reliability profiles $\{\rho_{1}^{m},\ldots,\rho_{M}^{m},\rho_{1}^{u},\ldots,\rho_{N}^{u}\}$ of validators and users are updated as follows. For each validator $m_{k}$ denote $\mathcal{Q}_k(n) = \{ q_k (x_\ell, x_j) \in \{0,1\}:
x_\ell \in X(n), x_j \in W_\ell \}$ as the set of binary perceptions $m_k$ sent in epoch $n$. And for each pair $(x_{\ell}, x_{j})$, denote $q^*(x_\ell, x_j) \in \{0,1\}$ as the majority values of all perceptions $\{ q_k (x_\ell, x_j) \}$ sent by validators in epoch $n$. Then the reliability of validator $m_{k}$ is updated as:
\begin{equation}\label{update1}
\begin{split}
\rho_{k}^{m}(n)&=
\zeta_{1}\rho_{k}^{m}(n-1)\\&+\frac{1-\zeta_{1}}{\lvert \mathcal{Q}_k(n) \rvert}\sum\limits_{q_{k}(x_{\ell},x_{j})\in \mathcal{Q}_k(n)}q^{*}(x_{\ell},x_{j})\oplus q_{{k}}(x_{\ell},x_{j}).
\end{split}
\end{equation}
Moreover, let $X_s(n) \subset X(n)$ be the set of transactions submitted by user $s$. If $X_s(n) \neq \phi$, then the reliability of user $s$ is updated as: 
\begin{equation}\label{update2}
\rho_{s}^{u}(n)=\zeta_{2}\rho_{s}^{u}(n-1)+\frac{1-\zeta_{2}}{\lvert  X_s(n) \rvert}\sum\limits_{x_{\ell}\in X_s(n)}V^{\ell},
\end{equation}
Otherwise $\rho_{s}^{u}(n)=\rho_{s}^{u}(n-1)$. In (\ref{update1})-(\ref{update2}), $0<\zeta_{1},\zeta_{2}<1$ are forgetting factors to count for time evolution.\par
Furthermore, the weight vector $\mathbf{y}_{k}$ for each validator $m_{k}$ is updated every $\mathcal{N}$ epochs. Denote $\mathcal{X}(t)=\bigcup_{n=(t-1)\mathcal{N}+1}^{n=t\mathcal{N}} X(n)$ as the set of all transactions that are submitted in epochs $(t-1)\mathcal{N}+1, \ldots,t\mathcal{N}$.  Denote further $\mathcal{X}_{k}(t)\subset \mathcal{X}(t)$ such that for each $x_{\ell}\in \mathcal{X}_{k}(t)$, $m_{k}\in \mathcal{V}_{\ell}$. Then the weight vector $\mathbf{y}_{k}(t)$ is given by the following optimization problem:
\begin{equation}\label{optimizz}
\begin{split}
\mathbf{y}_{k}(t) = \operatorname*{arg\,max}_{\mathbf{y}_{k}} \Bigg( & -\frac{1}{\mathcal{N}}\sum_{x_{\ell}\in \mathcal{X}_{k}(t)}V^{\ell}\log(\sigma(\mathbf{a}_{\ell}^{T}\mathbf{y}_{k})) \\
& + (1-V^{\ell})\log(1-\sigma(\mathbf{a}_{\ell}^{T}\mathbf{y}_{k})) \Bigg)
\end{split}
\end{equation}

that will be used for epochs $t\mathcal{N}+1, \ldots,(t+1)\mathcal{N}$. The minimization problem in (\ref{optimizz}) involves the log loss function, which is commonly used for binary classification~\cite{EEE1}. In this context, $\mathbf{a}_{\ell}$ represents the feature vector, and $\mathbf{y}_{k}$ is the classifier. The log loss function quantifies the discrepancy between the predicted probabilities and the true labels, effectively penalizing incorrect predictions and promoting well-calibrated probability estimates. Once the classifier $\mathbf{y}_{k}$ has been obtained, it assigns higher scores to transaction $x_{\ell}$ (i.e., a larger value of $\sigma(\mathbf{a}_{\ell}^T\mathbf{y}_{k})$) if it resembles valid transactions, and lower scores if it resembles invalid transactions. Validator $m_{k}$ utilizes these scores to calibrate the thresholds in (\ref{thresholds}) for accepting or rejecting transactions in subsequent epochs. \par

\subsection{Updating Rules}
Recall that as shown in (\ref{thresholds}), the validator's decision to accept or reject $x_{\ell}$ is based on $p_{k}^{\ell}(r)$, which represents validator $m_{k}$'s \emph{actual belief} at the end of round $r$ about the validity of transaction $x_{\ell}\in X(n)$. In addition, we introduce $\psi_{k}^{\ell}(r)$, which represents the \emph{intermediate belief} of validator $m_{k}$ at the end of round $r$ regarding the validity of $x_{\ell}$, which is used to find the actual belief $p_{k}^{\ell}(r)$. The calculation $p_{k}^{\ell}(r)$ and $\psi_{k}^{\ell}(r)$ are explained next.\par

\subsubsection{Updating actual beliefs}
In the literature on distributed consensus algorithms~\cite{QQQ7001,QQQ7002,QQQ7003}, the actual belief $p_{k}^{\ell}(r)$ at each round $r$ can be expressed using the general form $p_{k}^{\ell}(r)=g\left( \left\{p_{k^{\prime}}^{\ell}(r-1)\right\}_{k^{\prime} \in \mathcal{M}\cup\{k\}},\psi_{k}^{\ell}(r) \right)$.
Here, $g$ is called the opinion pooling function that is algorithm-specific, and $\left\{p_{k^{\prime}}^{\ell}(r-1)\right\}_{k^{\prime} \in \mathcal{M}\cup\{k\}}$ is the set of all actual beliefs that validator $m_{k}$ receives from other validators $m_{k^\prime}$, in addition to its own actual belief. At each round $r$, each validator takes into account its intermediate belief and the actual beliefs of other validators from the previous round to determine the actual belief for the current round.\par 
Consensus algorithms, such as those in~\cite{QQQ7004,QQQ7005}, typically use an opinion pooling function $g$ that is a linear or log-linear averaging function. However,~\cite{QQQ7006} shows that the min-rule can be more effective in achieving faster convergence in learning the correct actual belief, and better Byzantine resilience. Therefore, in HybridChain, at each round $r$ of epoch $n$, each validator $m_{k}$ updates the actual belief $p_{k}^{\ell}(r-1)$ for each transaction $x_{\ell}\in X_{r}(n)$ 
as follows.\par
First, validator $m_{k}$ sends its actual belief $p_{k}^{\ell}(r-1)$ to all other validators $m_{k^\prime}\in \mathcal{V}_{\ell}$ and receives $p_{k^\prime}^{\ell}(r-1)$ from all other validators $m_{k^\prime}$. Next, it sorts the received values $p_{k^\prime}^{\ell}(r-1)$ and removes the $f$-highest and $f$-lowest values. The set of remaining validators whose actual beliefs are not discarded by validator $m_{k}$ at round $r$ is denoted by $\mathcal{D}_{k}^{\ell}(r)$. The actual belief $p_{k}^{\ell}(r)$ is then computed as
\begin{equation}\label{rule2}
p_{k}^{\ell}(r)=\min \left\{\left\{p_{k^{\prime}}^{\ell}(r-1)\right\}_{k^{\prime} \in \mathcal{D}_{k}^{\ell}(r)\cup \{k\}}, \psi_{k}^{\ell}(r)\right\}. 
\end{equation}\par
\emph{Remark 5:} Suppose $W_{\ell}^{k}(r)=x_{j}$ and that validator $m_{k}\in \mathcal{V}_{\ell}$ and $m_{k}\in \mathcal{V}_{j}$. Then in case $q_{{k}}(x_{\ell},x_{j})=0$, instead of updating $p_{k}^{\ell}(r)$ in Stage 2, validator $m_{k}$ sets $p_{k}^{\ell}(r)=0$ and follows the rest of the processing procedure as usual.\par
\emph{Remark 6:} The rationale behind selecting $\lambda=\lfloor M/(2f+2) \rfloor$ is to ensure that, in (\ref{rule2}), after removing the $f$-highest and $f$-lowest values of $p_{k^\prime}^{\ell}(r-1)$ by validator $m_k$, at least one validator's opinion other than $m_k$ remains. This condition can be satisfied by having a minimum of $2f+2$ validators store each transaction, so that in the worst case even if both $m_k\in\mathcal{V}_{\ell}$ and $m_k\in\mathcal{V}_{j}$ for some $x_{j}\in W_{\ell}$, at least one other validator's opinion is incorporated.\par

\subsubsection{Updating intermediate beliefs}
To determine $p_{k}^{\ell}(r)$, as shown in (\ref{rule2}), validator $m_{k}\in \mathcal{V}_{\ell}$ must first obtain its intermediate belief $\psi_{k}^{\ell}(r)$. As previously stated, $\psi_{k}^{\ell}(r)$ is computed based on the perceptions $q_{k^\prime}$ that other validators $m_{k^\prime}$ send to validator $m_{k}$. To be more specific, let $Q_{k}^{\ell}(r)$ denote the set of all perceptions $q_{k^\prime}(x_{\ell},W_{\ell}^{k^\prime}(r))$ received by validator $m_{k}$ during Step 2 of Stage 1. To determine $\psi_{k}^{\ell}(r)$ in round $r$, validator $m_{k}$ applies the Bayesian rule as in (\ref{rule1}).
%\begin{strip}
%\begin{equation}\label{rule1}
%\psi_{k}^{\ell}(r)=\frac{P\left(Q_{k}^{\ell}(r) \mid x_{\ell}\,\, \text{is valid}\right) \psi_{k}^{\ell}(r-1)}{P\left(Q_{k}^{\ell}(r) \mid x_{\ell}\,\, \text{is valid}\right) \psi_{k}%^{\ell}(r-1)+P\left(Q_{k}^{\ell}(r) \mid x_{\ell}\,\, \text{is invalid}\right)(1-\psi_{k}^{\ell}(r-1))}.
%\end{equation}
%\end{strip}

\begin{figure*}[b]
    \begin{equation}\label{rule1}
    \psi_{k}^{\ell}(r)=\frac{P\left(Q_{k}^{\ell}(r) \mid x_{\ell}\,\, \text{is valid}\right) \psi_{k}^{\ell}(r-1)}{P\left(Q_{k}^{\ell}(r) \mid x_{\ell}\,\, \text{is valid}\right) \psi_{k}^{\ell}(r-1)+P\left(Q_{k}^{\ell}(r) \mid x_{\ell}\,\, \text{is invalid}\right)(1-\psi_{k}^{\ell}(r-1))}.
    \end{equation}
\end{figure*}

The conditional probabilities in (\ref{rule1}) are computed using the reliabilities $\rho^{m}_{k^\prime}$ of the validators $m_{k^\prime}$ who contributed the perceptions $q_{m_{k^{\prime}}}$ in $Q_{k}^{\ell}(r)$ as follows:
\begin{subequations}
\begin{equation}\label{eq:sub1}
\begin{split}
P\left(Q_{k}^{\ell}(r) \mid x_{\ell}\,\, \text{is valid}\right) &= \\
&\hspace{-5.5em}\prod_{\substack{q_{{k^{\prime}}}}\in Q_{k}^{\ell}(r)}\biggl[\rho^{m}_{k^\prime}\mathbf{1}_{[q_{{k^{\prime}}}=1]}+(1-\rho^{m}_{k^\prime})\mathbf{1}_{[q_{{k^{\prime}}}=0]}\biggl],
\end{split}
\end{equation}
\begin{equation}\label{eq:sub2}
\begin{split}
P\left(Q_{k}^{\ell}(r) \mid x_{\ell}\,\, \text{is invalid}\right) &= \\
&\hspace{-7em}\prod_{\substack{q_{{k^{\prime}}}}\in Q_{k}^{\ell}(r)}\biggl[((1-\beta) \rho^{m}_{k^\prime}+\beta(1-\rho^{m}_{k^\prime}))\mathbf{1}_{[q_{{k^{\prime}}}=1]} \\
&\hspace{-4em}+(\beta\rho^{m}_{k^\prime}+(1-\beta)(1-\rho^{m}_{k^\prime}))\mathbf{1}_{[q_{{k^{\prime}}}=0]}\biggl],
\end{split}
\end{equation}
\end{subequations}
where $\beta=\frac{2^{\lvert W_{\ell}-1 \rvert}}{2^{\lvert W_{\ell} \rvert}-1}$ is the probability of $x_{\ell}$ being in conflict with a given $x_{j}\in W_{\ell}$ given that $x_{\ell}$ is an invalid transaction. The denominator of $\beta$ is the total number of possible perceptions of the conflict of $x_{\ell}$ with respect to transactions in $W_{\ell}$, excluding the case that it is not in conflict with any in $W_{\ell}$; the numerator of $\beta$ is the total number of perceptions given $x_{\ell}$ is in conflict with $x_{j}$.\par
Finally, it should be mentioned that validators use $p_{k}^{\ell}(0)=\rho_{s}^{u}$ and $\psi_{k}^{\ell}(0)=\rho_{s}^{u}$ as initial values in (\ref{rule2}) and (\ref{rule1}), where $\rho_{s}^{u}$ is the reliability of user $s$ that sends transaction $x_{\ell}$.\par

\section{SECURITY ANALYSIS AND COMPARISON}
In this and next sections, we will conduct a comparative analysis between HybridChain and two other notable schemes in the field of distributed ledgers: IOTA, a benchmark DAG scheme, and Omniledger, a sharded blockchain. The primary objective of this section is to evaluate  security features of all three schemes.
\subsection{Replay Attack} A replay attack~\cite{EEE2} occurs when an attacker intercepts a transaction and resends it at a later time, in an attempt to deceive the recipient into thinking that a new transaction is taking place, when in fact it is just a repeat of a previous transaction. Replay attacks can be particularly problematic in multi-party transactions, where all parties must agree to commit to a certain action simultaneously, or else the entire transaction is rolled back. In this scenario, a replay attack could cause incorrect decisions to be made about the transaction, leading to fraud, loss of funds, or other security breaches.\par
IOTA Tangle is resilient against replay attacks. Transactions in IOTA Tangle are interlinked, forming a ``tangle'' that undergoes validation by network nodes, which ensures the uniqueness of each transaction and prohibits replication. Furthermore, every transaction in Tangle requires a distinct identifier called a ``branch'' and  a ``trunk,'' pointing to two previous transactions within the Tangle. When a transaction is broadcast to the IOTA network, it undergoes validation by the network nodes. If a node receives a transaction with a branch and trunk that have already been utilized, the transaction is rejected, effectively thwarting any replay attempts.\par
The Omniledger is vulnerable to replay attacks~\cite{QQQ571}, as it utilizes Atomic commitment for transaction processing. A potential attacker can easily execute a replay attack by intercepting a cross-shard transaction, and then resending it to the involved shards. Nodes within the network are unable to differentiate between honest but delayed transactions and maliciously replayed ones, resulting in conflicting views among different shards regarding the transactions in question. This is particularly concerning for cross-shard transactions, as validators in each shard lock the transactions until a decision is made to either commit or abort the transaction. In the case of a replay attack, the transaction may become locked indefinitely, or even result in a double-spending scenario.\par 
In contrast to Omniledger, HybridChain is resilient against replay attacks due to its unique design. While similar to sharded blockchains in the sense that our scheme processes transactions in parallel, it is not a sharded blockchain itself. In contrast to sharded systems where only a subset of nodes process transactions, non-sharded blockchains require consensus from all nodes for each transaction, ensuring that it is unique and cannot be replayed at a later time. Specifically, for each transaction $x_{\ell}\in X(n)$, the set of validators in $\mathcal{V}_{j}$ for each transaction $x_{j}\in W_{\ell}$ are randomly assigned and can be entirely different. This means that each validator $m_{k}\in \mathcal{V}_{\ell}$ receives a set of perceptions $Q_{k}^{\ell}(r)$ from all other validators in the network, making each validator a participant in the verification of transaction $x_{\ell}$.\par
\subsection{Message Withholding Attack} A message withholding attack~\cite{EEE3} involves a malicious actor deliberately withholding important messages or transactions that are necessary for the system to function correctly. In particular, the attacker can concentrate on a particular transaction or set of transactions, and prevent the circulation of messages required to authenticate them. As a result, the transaction may not be included in a block, or its confirmation could be postponed, causing disruptions in the normal functioning of the system. Message withholding attacks are particularly potent in systems that employ a consensus mechanism based on a quorum to approve transactions. Through the suppression of messages, the attacker can impede the system's ability to reach consensus and even cause a fork in the blockchain or ledger.\par
In IOTA's Tangle, the act of withholding a single message does not necessarily hinder the confirmation of a transaction since the confirmation process does not depend solely on a linear sequence of transactions. Rather, transactions are verified based on a consensus mechanism that considers the entire network of interconnected transactions. This distinctive architecture and consensus mechanism of IOTA make it resilient to message withholding attacks.\par
On the other hand, Omniledger is susceptible to the message withholding attack. Specifically, for intra-shard transactions, the shard leader, responsible for processing the transaction, receives the transaction and dispatches it to the validators within the shard. If the shard leader is malicious and launches a message withholding attack, it can prevent the transaction from being processed. Also, in the case of cross-shard transactions, an adversary can create a cross-shard transaction and send the transaction to the source shard but withhold the transaction from the destination shard causing the cross-shard transaction to remain active and keeping the validators occupied with the transaction~\cite{QQQ571}.\par
HybridChain is resilient to message withholding attacks. To process a transaction $x_{\ell}$, each of the witness transactions $x_{j}\in W_{\ell}$ is stored by $2f+2$ validators, assuming that the network has no more than $f$ dishonest validators. Validator $m_{k}\in \mathcal{V}_{\ell}$ receives perceptions $q_{m_{k^\prime}}(x_{\ell},x_{j})$ from validators $m_{k^\prime}\in \mathcal{V}_{j}$ for all transactions $x_{j}\in W_{\ell}$. Hence, in the event of a message withholding attack by dishonest validators, at least $f+2$ other validators exist to inform validator $m_{k}$ of the transaction's validity with respect to each witness transaction.\par
\subsection{Orphanage Attack} In a distributed ledger scheme, when a conflict occurs between a new block or transaction and an existing one, the network must choose to orphan one of them. If the scheme employs Nakamoto consensus, where the validity of blocks or transactions is determined by the ledger structure, selecting the new block or transaction results in the old one being orphaned, wasting the efforts put into its creation. This problem becomes more severe when a subsequent block or transaction conflicts with the chosen block or transaction. If the network selects the subsequent block or transaction, it not only orphanes the old block or transaction but also orphanes the one it was chosen over. \par
The orphanage attack ~\cite{QQ22} involves taking advantage of the orphanage problem and create transactions or blocks to make the existing ones to be left behind. By conducting simulations and analyzing particular strategies,~\cite{QQ22} revealed that the impact of the orphanage attack could be significant in IOTA, to the extent that a sufficiently powerful attacker could cause $96\%$ of transactions in IOTA to be orphaned.\par
On the other hand, for schemes that do not employ Nakamoto consensus, if a conflict arises between a new transaction or block and an existing one, the network keeps the existing one and discards the new one. The Omniledger uses Fast Byzantine Consensus (FBC) as the consensus algorithm, which is a classical consensus. FBC can resist orphanage attacks by providing immediate finality for confirmed blocks, employing a synchronized confirmation process to prevent manipulation. These features ensure that once a block is confirmed by the network, it cannot be discarded or orphaned.\par
Similar to Omniledger, in HybridChain all submitted transactions are placed in a queue and processed in batches in parallel during epochs, ensuring that none of the transactions are orphaned due to acceptance of another transaction. Thus, HybridChain is immune to the orphanage attack.\par 
\subsection{Routing Attack} The Border Gateway Protocol (BGP) is a protocol used by Autonomous Systems (AS), which are groups of interconnected Internet Protocol (IP) routing prefixes controlled by one or more network operators, to share information about subnet reachability through advertisements. BGP does not verify the accuracy of advertised routes, making it possible for a malicious AS to add and propagate advertisements for routes that do not actually exist. BGP hijacking occurs when an AS advertises a route that it does not have in order to attract Internet traffic intended for that destination.\par
%%%%%%%%%%%%%%%%%%%%%%%%%%%%%%%%%%%%%%%%%%%%%%%%%%%%%%%%%%%%%%%%%%%%%%%%%%%%%%%%%%%%%%
In IOTA, a full node must connect with random full nodes to maintain up-to-date information. This randomness is critical to prevent adversaries from compelling specific nodes to become peers, and to prevent predictable network configurations that could result in eclipse attacks. In an eclipse attack, a group of colluding nodes surrounds a target node and intercepts all communication with the rest of the network, potentially causing ledger consensus issues such as double spending or forking. In~\cite{QQ23} it is demonstrated that BGP hijacking may be successful in the IOTA network unless two full nodes are in the same AS and use intra-AS routes. However, forcing nodes to be neighbors in the same AS would reduce the network's overall connectivity, which is undesirable.\par
%%%%%%%%%%%%%%%%%%%%%%%%%%%%%%%%%%%%%%%%%%%%%%%%%%%%%%%%%%%%%%%%%%%%%%%%%%%%%%%%%%%%
On the other hand, in contrast to IOTA, Omniledger is immune to such routing attack. This is because in Omniledger, all nodes hold a partial copy of the ledger, and the processing of transactions is distributed among all nodes. To prevent routing attacks, OmniLedger uses a hybrid routing approach that combines both centralized and decentralized routing mechanisms. In this approach, the routing information is divided into two parts: a centralized directory that contains the IP addresses and public keys of all the shards, and a decentralized routing table that contains the routing information for each shard. The centralized directory helps to prevent attacks on the routing information, while the decentralized routing table helps to prevent attacks on the transaction history of each shard.\par
%%%%%%%%%%%%%%%%%%%%%%%%%%%%%%%%%%%%%%%%%%%%%%%%%%%%%%%%%%%%%%%%%%%%%%%%%%%%%%%%%%%%
Moreover, HybridChain achieves a well-distributed computation of transaction processing. Similar to Omniledger, it avoids relying on centralized entities such as full nodes. Instead it allows all validators to access the necessary information by distributing the ledger data. Thus, the routing attack cannot compromise the consistency or security of the ledger data. In particular, even if the attacker
 targets validator $m_{k}$, the decision $v_{k}^{\ell}$ and the final decision $V^{\ell}$ regarding transaction $x_{\ell}$ are unlikely to be affected. This is because the calculation of $p_{k}^{\ell}(r)$, which determines $v_{k}^{\ell}$, depends on all perceptions in $Q_{k}^{\ell}(r)$ for each round $r$, where $Q_{k}^{\ell}(r)$ is a group of perceptions obtained from various validators. Therefore, in order to change $v_{k}^{\ell}$, the attacker must disrupt many connections between validator $m_{k}$ and other validators, making it difficult to influence the decision-making process of any validator, and consequently the final decision.\par
%%%%%%%%%%%%%%%%%%%%%%%%%%%%%%%%%%%%%%%%%%%%%%%%%%%%%%%%%%%%%%%%%%%%%%%%%%%%%%%%%%%%%
Finally, we summarize whether or not each distributed ledger scheme is resistant to various attacks in Table \ref{attack}.

\begin{table}[htb]
\centering
\hspace*{-0.5cm}\resizebox{0.45\textwidth}{!}{\begin{tabular}{|c"c|c|c|c|c|}
\hline
\textbf{Attack Type} & \textbf{IOTA} & \textbf{Omniledger} &\textbf{HybridChain} \\ \thickhline
\textbf{Replay Attack} & Yes & No & Yes\\   \hline
\textbf{Message Withholding Attack} & Yes & No & Yes \\  \hline
\textbf{Orphanage Attack} & No & Yes & Yes\\    \hline
\textbf{Routing Attack} & No & Yes & Yes\\      \hline
\end{tabular}}
\caption{Resistance to various attacks by different distributed ledger schemes.}
\label{attack}
\end{table}

%%%%%%%%%%%%%%%%%%%%%%%%%%%%%%%%%%%%%%%%%%%%%%%%%%%%%%%%%%%%%%%%%%%%%%%%%%%%%%%%%%%%%%

\section{Simulation Results}
This section provides a thorough comparison of HybridChain with IOTA and Omniledger, focusing on the system's latency, throughput, and scalability, to showcase the advantages of HybridChain. Additionally, we assess how our decentralized scheme compares with centralized classification algorithms, that are commonly used for detecting fraudulent transactions.\par
\subsection{Simulation Setups}
\subsubsection{IOTA}
To simulate IOTA, we utilized the simulator TangleSim~\cite{QQ26}  along with certain software modifications and measurement tools found in ~\cite{QQ27} and~\cite{QQ28}, which simulate an improved version of the IOTA~\cite{QQQ2001}, where the nodes in the network use the Approval Weight (AW) mechanism to reach consensus about the set of the valid transactions.\par
Specifically, when a light node intends to attach a transaction to the Tangle it must select $\mathcal{P}\geq 2$ older transactions using the Restricted Uniform Random Tip Selection (R-URTS) protocol, which chooses parents uniformly at random from the pool of tip transactions whose age are less than a threshold $\delta$. Then to reach consensus about the set of valid transactions, nodes first calculate the \emph{Mana} score for all participant nodes based on their ownership of a scarce resource, which represents their influence. Then using the Mana score, nodes determine the set of valid transaction in the Tangle by computing the \emph{cumulative weight} of all transactions and comparing it with a predefined threshold $\theta$. If the cumulative weight of a transaction exceeds $\theta$, it is considered valid.\par
In the IOTA simulations, honest nodes select parents following the R-URTS protocol and issue valid transactions. The adversary nodes always issue invalid transactions; and while choosing a parent they select the one with the highest possible age; and if possible they try to select their own transactions.\par    
Denote the rate of transaction issuance per minute by $\gamma=\gamma_{a}+\gamma_{h}$, where $\gamma_{a}$ and $\gamma_{h}$ are transaction issuance rates of adversary nodes and honest nodes, respectively. Define $S=\frac{\gamma_{a}}{\gamma}$ as the spamming rate of the adversary nodes. Each transaction must select $\mathcal{P}=2$ other transactions as parents with an age restriction of $\delta=1$ minute. Also the confirmation threshold $\theta=0.66(1-\tau)$ where $\tau$ is the percentage of adversary nodes. We set the number of nodes in IOTA as $M=1000$. In each simulation, we execute the software for a duration of $5$ minutes, generating $5\gamma$ transactions at various transaction issuance rates denoted by $\gamma$. These transactions are then attached to the Tangle, forming an integral part of the simulation process.\par
\subsubsection{Omniledger}
To simulate Omniledger, we utilized OverSim~\cite{QQQ102}, a modular framework, on OMNeT++ 6.0~\cite{QQQ100}, a process-based discrete-event simulator. The simulation parameters are set based on ~\cite{QQQ2000} as follows. The required percentage of votes for accepting a transaction, as well as other consensus-related activities like leader selection, is set to $67\%$. Additionally, the block size is fixed at $1$ MB. The task completion and voting timeout values are set to $500$ milliseconds, and an epoch transition occurres when all tasks are completed, and the network has processed the entire task list. For all simulations, the bandwidth of communication links and latency between validators are set to $20$ Mbps, and $100$ ms, respectively.\par
To generate transactions, we employed Bitcoin Core~\cite{QQQ5000} version 23.1, which was the latest release at the time of writing this paper. The rate of incoming transactions in Omniledger is represented by $\gamma$, and the total number of generated transactions in all simulations matched that in IOTA, which is $5\gamma$.\par
In Omniledger, it is important to note that intra-shard and cross-shard transactions are processed separately, with cross-shard transactions generally taking longer to process than intra-shard ones. The number of shards in Omniledger, denoted as $\Gamma$, varied across different simulations. We provide the specific value of $\Gamma$ for each simulation. Based on~\cite{QQQ6000}, using the RandHound algorithm in Omniledger with $\Gamma=4$, approximately $81\%$ of transactions are categorized as cross-shard, while with $\Gamma=16$, nearly $95\%$ of transactions are classified as cross-shard. Thus, the performance of Omniledger is expected to be heavily influenced by the value of $\Gamma$.\par
In our simulations, we considered $M=1000$ validators in Omniledger. The percentage of Byzantine dishonest validators, denoted as $\tau$, was also specified for each simulation, similar to IOTA. A Byzantine dishonest validator adheres to the network's protocol but intentionally casts incorrect votes for both intra-shard and cross-shard transactions, whereas an honest validator always votes correctly.\par

\subsubsection{HybridChain}
To simulate HybridChain, which shares similarities with Omniledger, we employed OverSim as our simulation platform. However, to customize and implement our scheme, we made significant modifications to four key modules of Omniledger.\par
Firstly, we modified the validator module, which represents a validator in the blockchain network. Our modification accounted for the random grouping of validators in each epoch and included operations at all of Stages described in Section III-C, and managing validators' interactions during each round.\par
Secondly, we made changes to the consensus module responsible for achieving consensus among validators regarding the set of valid transactions. These modifications encompassed adjusting the voting and decision-making logic to incorporate HybridChain's consensus mechanism detailed in Section III.\par
Thirdly, we modified the epoch module, which represents the epochs in HybridChain comprising multiple rounds. We redefined the structure and duration of each epoch, as well as the transitions between epochs. Similar to Omniledger, we set the task completion and voting timeout values to $500$ milliseconds, with a round transition occurring when all tasks are completed, and the network has processed the entire task list. We also set the bandwidth $20$ Mbps and imposed a latency of $100$ ms on all communication links in all simulations.\par
Lastly, we adapted the transaction module to accommodate the specific transaction structure and validation rules of HybridChain. Specifically, in each simulation, similar to the corresponding simulations for IOTA and Omniledger, we generated the same number of $5\gamma$ transactions, consisting of both valid and invalid transactions. Additionally, for all simulations, we utilized the probability distributions listed in Table \ref{distributions} to generate the attribute vector $\mathbf{a}_{\ell}$ for valid or invalid transactions $x_{\ell}$. Here, it should be noted that the values of the elements in the attribute vector are truncated, to certain ranges as indicated in the last column.\par
\begin{table}[htb]
\centering
\hspace*{-0.5cm}\resizebox{0.45\textwidth}{!}{\begin{tabular}{|c"c|c|c|c|c|}
\hline
\textbf{Attribute Vector Element} & \textbf{Valid} & \textbf{Invalid}  & \textbf{Range}\\ \thickhline
${a}_{\ell}[1]$ & ${\rm Gamma}(3,1)$ & ${\rm Gamma}(17.5,0.5)$&$(0,10]$ \\   \hline
${a}_{\ell}[2]$& ${\rm Norm}(0.5,0.15)$ & ${\rm Norm}(0.25,0.075)$&$(0,+\infty)$ \\  \hline
${a}_{\ell}[3]$ & ${\rm Exp}(10)$  & ${\rm Exp}(5)$&$(0,50]$\tablefootnote{The generated number is rounded down to find a discrete number.}  \\   \hline
${a}_{\ell}[4]$ & ${\rm Pois}(1.5)$ & ${\rm Pois}(2.5)$&$[1,+\infty)$ \\\hline
${a}_{\ell}[5]$ & ${\rm Norm}(0.7,0.1)$ & ${\rm Norm}(0.4,0.1)$ & $(0,1]$\\\hline
\end{tabular}}
\caption{Distribution of attribute vector elements for generating valid and invalid transactions.}
\label{distributions}
\end{table}
To simulate an invalid transactions $x_{\ell}$, after determining the size $W_{\ell}$ using the distribution of ${a}_{\ell}[4]$ from Table \ref{distributions}, we randomly select transactions $x_{j}\in W_{\ell}$ to be in conflict with $x_{\ell}$.\par
Also, in (\ref{thresholds}) we use the sigmoid function $\sigma(\mathbf{a}_{\ell}^{T}\mathbf{y}_{k})=\frac{1}{1+e^{-\mathbf{a}_{\ell}^{T}\mathbf{y}_{k}}}$ and set $\mu_{1}=1$ and $\mu_{2}=0.5$, which results in $0.5\leq\eta_{1}(\mathbf{a}_{\ell},\mathbf{y}_{k})\leq1$ and $0\leq\eta_{2}(\mathbf{a}_{\ell},\mathbf{y}_{k})\leq0.5$. Moreover, we set the number of users $N=100$ and updating frequency $\mathcal{N}=20$ epochs. Futhermore, for the forgetting factors in (\ref{update1})-(\ref{update2}), we set $\zeta_{1}=0.98$, $\zeta_{2}=0.9$. Additionally, the reliability of validators and users are uniformly generated in range $[0.3,0.8]$.\par
To initially train the weight vector $\mathbf{y}_{k}$, we use $10000$ transactions, with half labeled as valid and half as invalid. We utilize the Scikit-learn~\cite{QQ00000} library in Python to solve the minimization problem in (\ref{optimizz}).\par
Similar to IOTA and Omniledger, the number of validators is set as $M=1000$, the percentage of dishonest validators is denoted by $\tau$. In HybridChain, honest validators always send correct perceptions to other validators and make local decisions following the defined rules in Section III. On the other hand, dishonest validators consistently send incorrect perceptions to other validators and make incorrect local decisions.\par
\subsection{Results}
\subsubsection{Classification}
In order to evaluate the performance of our decentralized consensus scheme in classifying the validity of transactions, we compared it to a centralized classification algorithm, the Random Forest algorithm~\cite{QQ25}, known for its high accuracy. To conduct the simulation, we generated a dataset of $10000$ transactions, which were used to train a Random Forest classifier consisting of 100 trees. Each tree utilized a random subset of two or three attributes from the five features in the attribute vector. Then, we evaluated the performance of the trained Random Forest model on a separate set of $5000$ transactions, comprising an equal number of valid and invalid transactions, yielding an accuracy of $98.2\%$.\par
\begin{figure}[!t]
\centering   
                \includegraphics[width=\columnwidth]{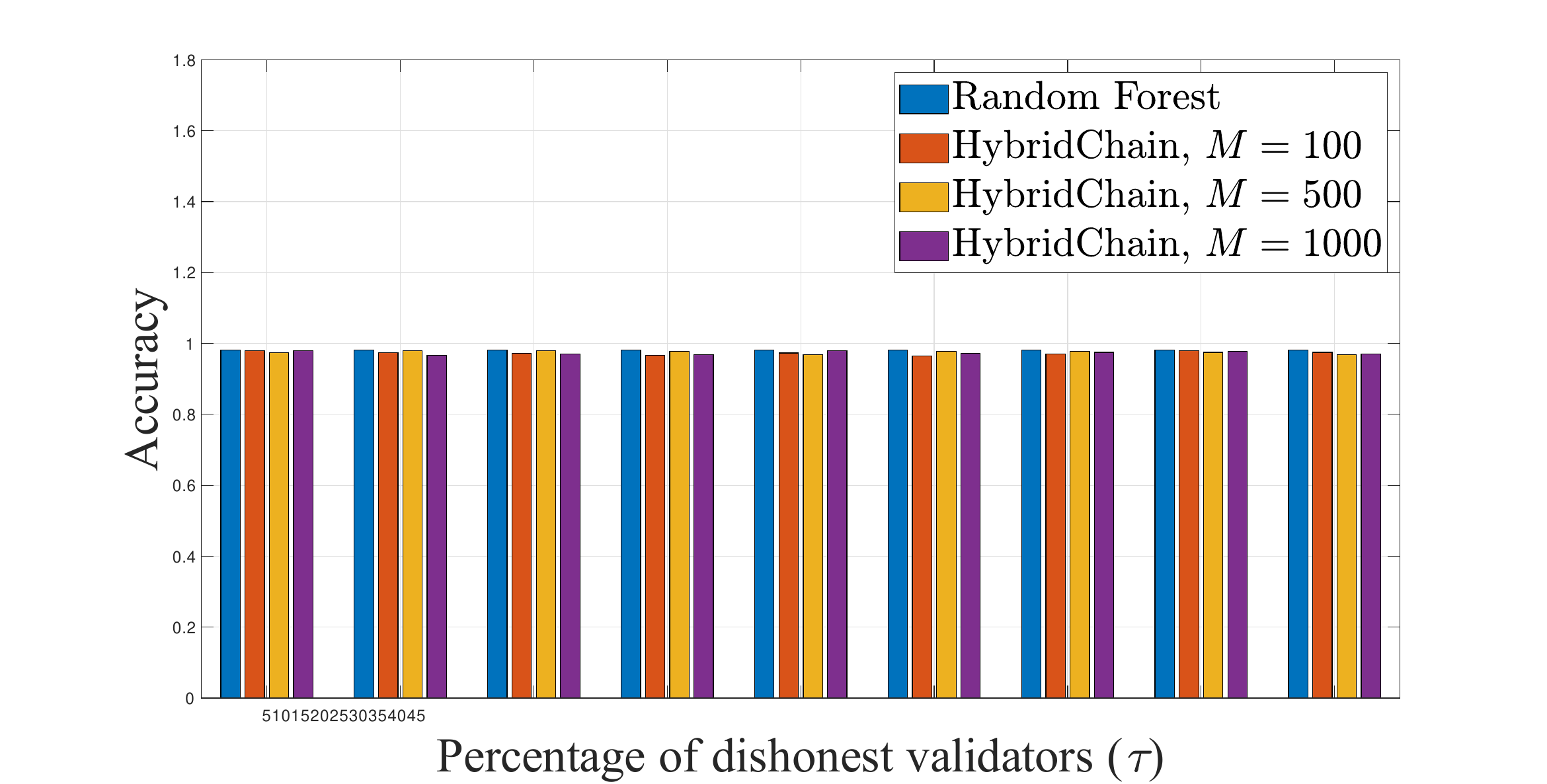}
                \label{CLTFig3}
\DeclareGraphicsExtensions.
\caption{Accuracy of transaction validity classification under different values of dishonest validators $(\tau)$ and number of validators $M$ for HybridChain compared to Random Forest.}
\label{forest}
\end{figure}

We subjected the same set of $5000$ transactions to HybridChain and a network with varying numbers of validators and different values of $\tau$. The results are depicted in Fig.\ref{forest}, demonstrating that HybridChain consistently exhibits stable performance, under different number of dishonest validators or different network sizes. This remarkable performance can be attributed to the collaborative decision-making mechanism of HybridChain, which aids validators in enhancing the accuracy of their decisions.\par
\subsubsection{Latency}
We define latency the amount of time it takes for a transaction to be confirmed and added to the ledger. It is the delay between the moment a transaction is initiated and the moment it is considered final and irrevocable.\par
\begin{figure}[!t]
\centering   
                \includegraphics[width=\columnwidth]{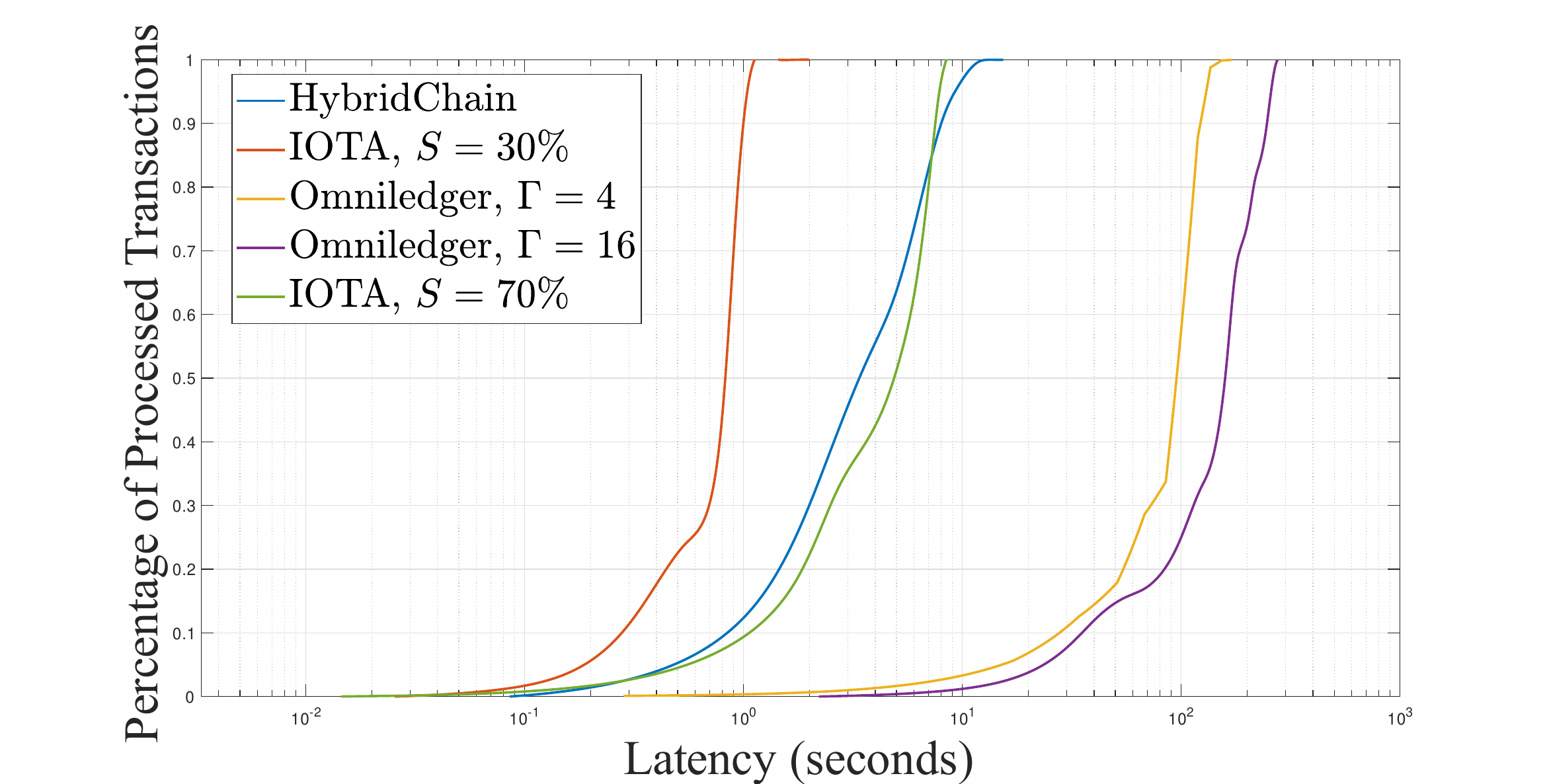}
                \label{CLTFig3}
\DeclareGraphicsExtensions.
\caption{Latency distribution with incoming transactions rate $\gamma=3\times 10^4$, $M=1000$ validators, and $\tau=20\%$.}
\label{LLL}
\end{figure}

Fig.\ref{LLL} presents the cumulative distributions of latencies of different systems with $\tau=20\%$ dishonest validators. For IOTA we set the spamming rates to $S=30\%$ and $S=70\%$ and set $\gamma=3\times 10^4$. For Omniledger we set $\Gamma=4$ and $\Gamma=16$ shards. The results demonstrate that Omniledger has the highest latency for both sharding sizes and IOTA with a spamming rate of $S=30\%$ has the lowest latency. The latency of HybridChain is slightly lower than that of IOTA with a spamming rate of $S=70\%$. However, as illustrated in Fig.\ref{SSS1} in Section V-B-4, the quick processing of IOTA with a spamming rate of $S=30\%$ comes at the expense of reduced verification accuracy.\par
\subsubsection{Throughput}
The throughput is the average number of transactions processed by a scheme per unit of time. However, when comparing two schemes with equal throughput, the one that achieves higher transaction validation accuracy outperforms the other. Therefore, when evaluating the throughput, it is essential to consider two factors: accuracy, which indicates the correctness of decisions made during transaction processing, and the time required for processing transactions. To assess the throughputs of the three schemes, we conducted two simulations.\par
We set the incoming transaction rate $\gamma=6000$. The three schemes were simulated for various values of $\tau$, ranging from $5\%$ to $45\%$. IOTA was simulated with three spamming rates $S\in \{30\%,\, 50\%,\, 70\%\}$. Omniledger was simulated with three values of $\Gamma\in\{4,\, 8,\, 16\}$, representing the number of shards in the network. In IOTA, a transaction was considered processed when its cumulative weight exceeds $\theta$. In Omniledger and HybridChain, a transaction was considered processed when it was either accepted or rejected by the network consensus. A one-minute time window was used to process transactions for all three schemes, with the number of processed transactions and accuracy recorded.\par
\begin{figure}[!t]
\centering{
 \subfigure[t][ 
    Throughput
 ]{
                \centering
                \includegraphics[width=\columnwidth]{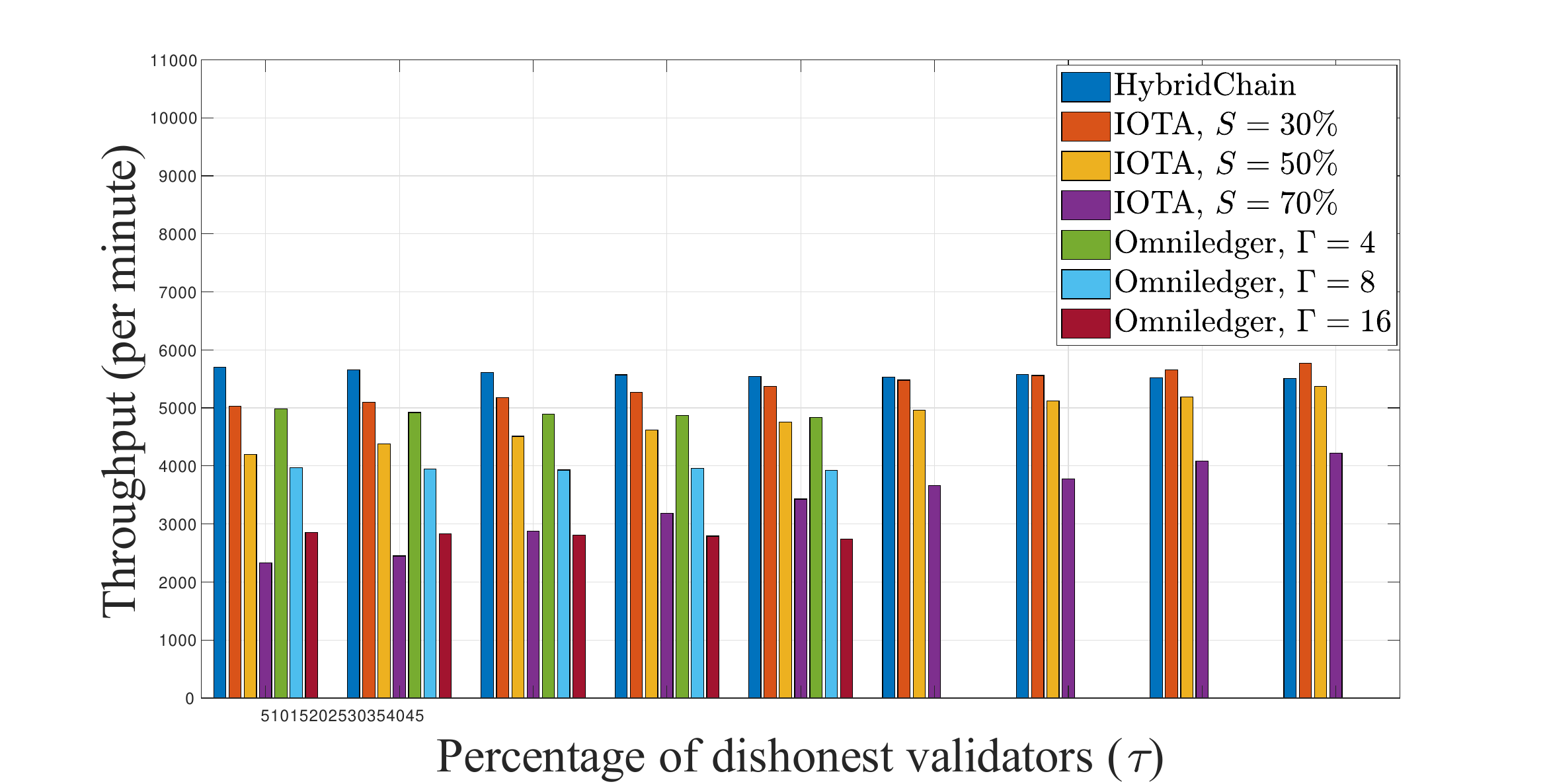}
                %\label{lowcantenna}
                }
               \subfigure[t][ 
    Accuracy
 ]{
                \centering
                \includegraphics[width=\columnwidth]{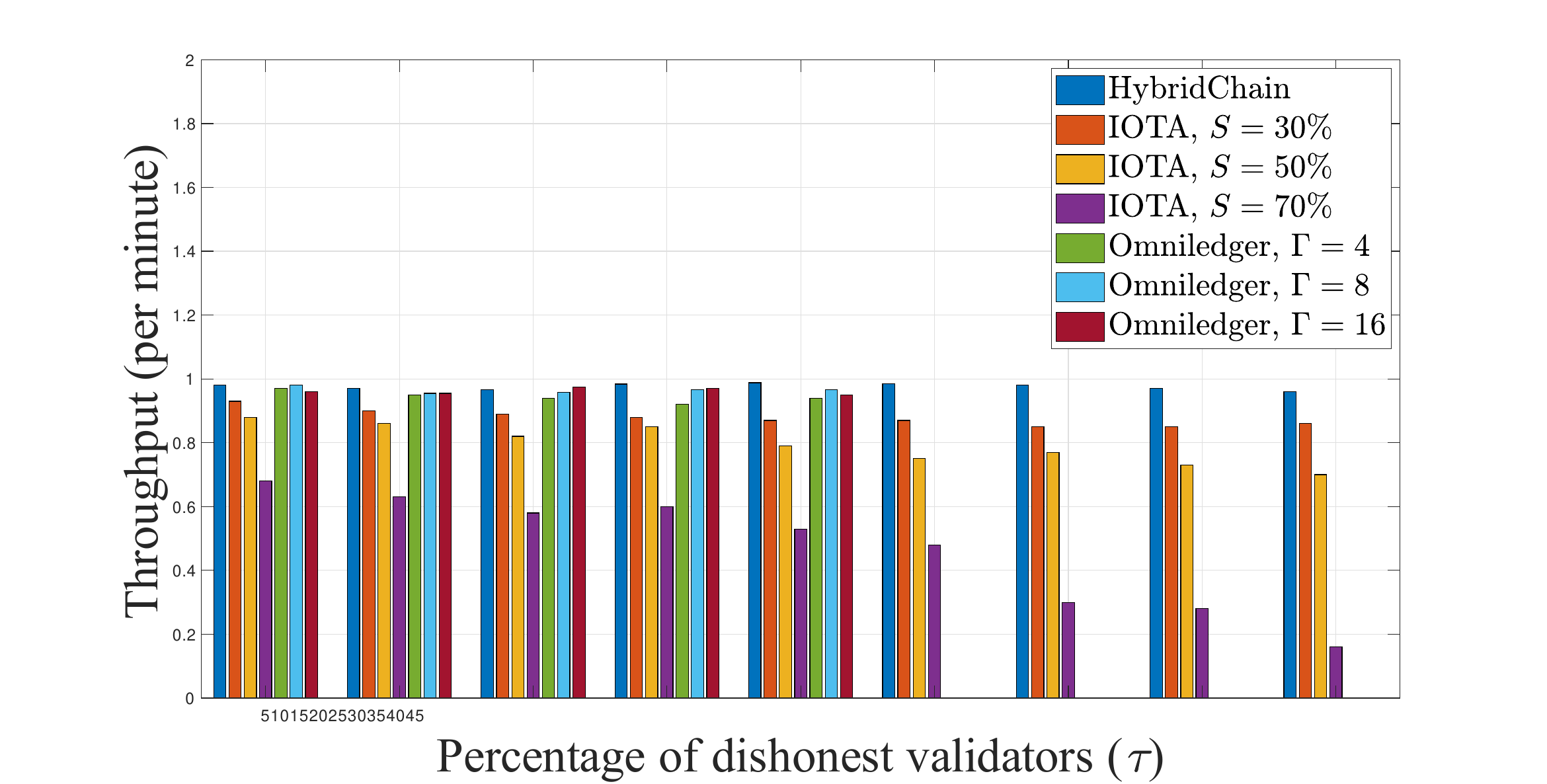}
                %\label{highcantenna}
                }     
                }
\DeclareGraphicsExtensions.
\caption{Throughput and accuracy based on percentage of dishonest validators, with incoming transactions rate $\gamma=6000$ and $M=1000$ validators.}
\label{TTT1}
\end{figure}

The results averaged over ten simulations runs, are shown in Fig.~\ref{TTT1}. It is evident that the spamming rate of adversaries heavily influences the throughput and accuracy of IOTA. Specifically, each adversary node impacts IOTA in two ways: first, by increasing the number of orphaned transactions, and second, by confirming invalid transactions. Consequently, when the spamming rate $S$ increases while keeping $\tau$ fixed, both the throughput and accuracy decrease. This outcome is anticipated, as stated in~\cite{QQ22}, due to the existence of a critical spamming rate denoted as $S_{critical}=\frac{\mathcal{P}-1}{\mathcal{P}}$. When $S$ exceeds $S_{critical}$, the inflation of the tip pool size accelerates, resulting in a higher number of orphaned transactions. Also, with an increase in the spamming rate $S$ of adversaries, more cumulative weights of invalid transactions surpass the threshold $\theta$. This occurs because adversary nodes, deviating from the R-URTS protocol, tend to select invalid transactions as parents for their submitted transactions, resulting in a decline in the accuracy of processed transactions.\par
Moreover, when $S$ is fixed, an increase in $\tau$ leads to an increase in throughput. This is because the threshold $\theta$ decreases as a result, allowing a larger number of transactions to be processed. However, the accuracy decreases since a larger number of invalid transactions are validated. The objective of setting the acceptance threshold $\theta$ as a function of $\tau$ is to simulate the Mana score in IOTA. Unlike Omniledger and HybridChain, where the number of adversary nodes affects the consensus results, IOTA's performance is unaffected by the quantity of adversary nodes. Instead, it is the distribution of the Mana scores of the nodes that influences the validity of transactions. Therefore, to facilitate a comparison between IOTA and the other two schemes for the same value of $\tau$, we consider $\theta$ as a function of $\tau$ to simulate the power of adversaries in the network.\par
For Omniledger, the results showed that for $5\%\leq\tau\leq25\%$, increasing $\Gamma$ from $8$ to $16$ led to reduced throughput. This reduction can be attributed to the increased number of cross-shard transactions, which subsequently prolongs the processing time. However, Omniledger demonstrated stable accuracy in its processed transactions. Moreover, with a fixed $\Gamma$, an increase in $\tau$ slightly decreased the throughput due to the additional communication rounds required to achieve consensus using the FBC protocol. The increase in the number of adversary validators increases the time required for consensus by the  FBC protocol. However, the accuracy of the processed transactions remained stable, thanks to Omniledger's design that involves validators in transaction processing. On the other hand, for $25\%\leq\tau\leq45\%$, the throughput dropped to almost zero for all $\Gamma$ values. This indicates Omniledger's inability to handle a network with more than $25\%$ dishonest validators.\par
As for HybridChain, its throughput is comparable to that of IOTA with $S=30\%$ and $\tau=45\%$. However, when $S=50\%$ and $S=70\%$, as well as in Omniledger for all values of $\Gamma$ and for all values of $\tau$, HybridChain performs significantly better. The reason is that HybridChain, like DAG schemes, accelerates transaction processing through partial verification. Moreover, as $\tau$ increases, the throughput of HybridChain slightly decreases since validators require more rounds in each epoch to reach consensus, thereby taking longer to process the transactions. Nevertheless, similar to the findings in Section V-B-1, an increase in $\tau$ does not impact the accuracy of HybridChain, as it distributes the relevant witness transactions for each transaction $x_{\ell}\in X(n)$ in an epoch among the validators $\mathcal{V}_{j}$ for transactions $x_{j}\in W_{\ell}$, in a manner that is impervious to the rise in the number of dishonest validators.\par
In a DAG distributed ledger system, reducing the volume of transactions can increase its susceptibility to attacks, as indicated by ~\cite{QQQ110}. To further investigate this phenomenon, we conducted a second simulation, keeping all settings identical to the first simulation but varying the rate of incoming transactions by setting $\gamma=1200$. This allowed us to observe the effects of low transaction volumes on throughput and accuracy.\par
\begin{figure}[!t]
\centering{
 \subfigure[t][ 
    Throughput
 ]{
                \centering
                \includegraphics[width=\columnwidth]{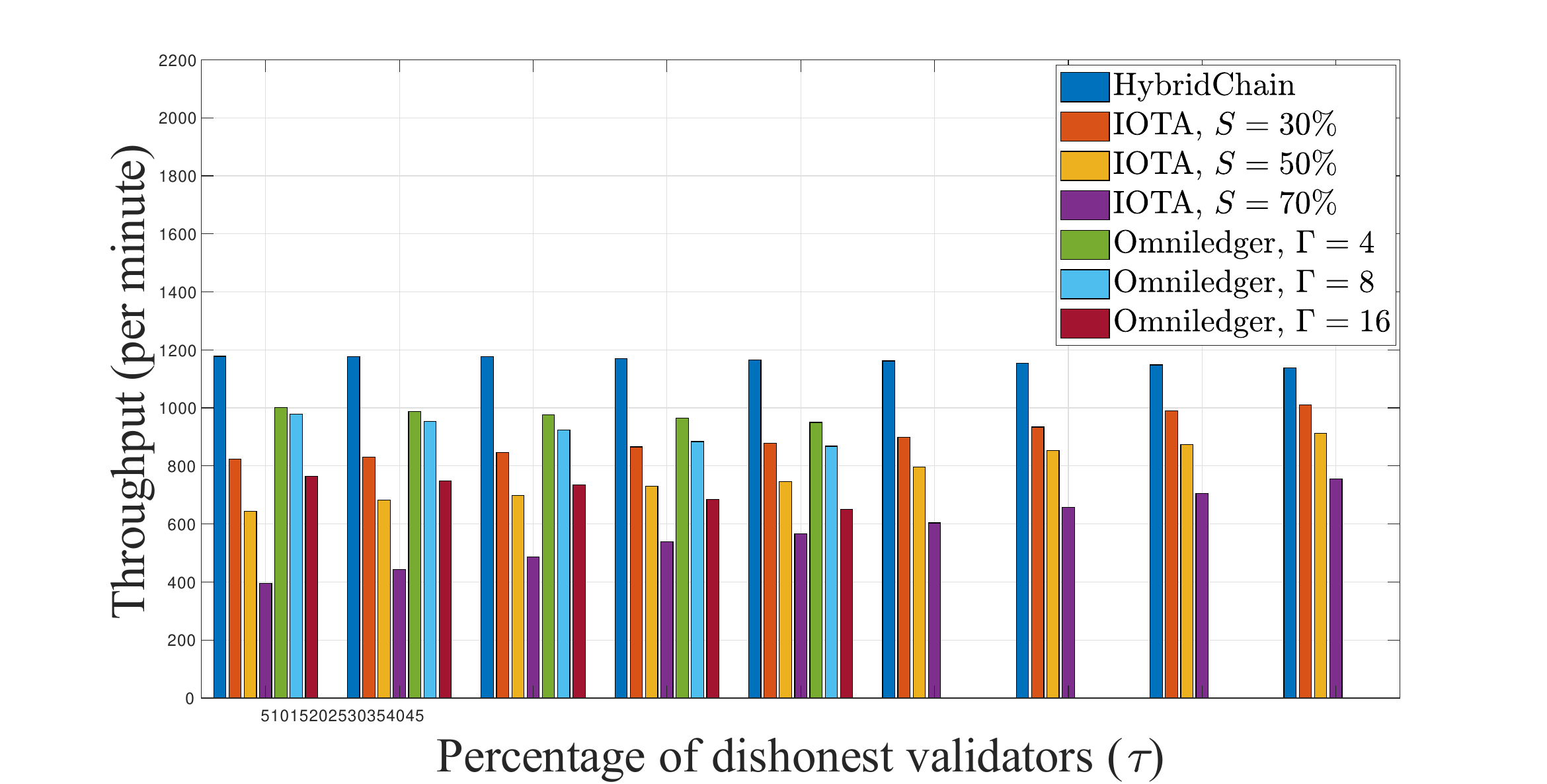}
                %\label{lowcantenna333}
                }
               \subfigure[t][ 
    Accuracy
 ]{
                \centering
                \includegraphics[width=\columnwidth]{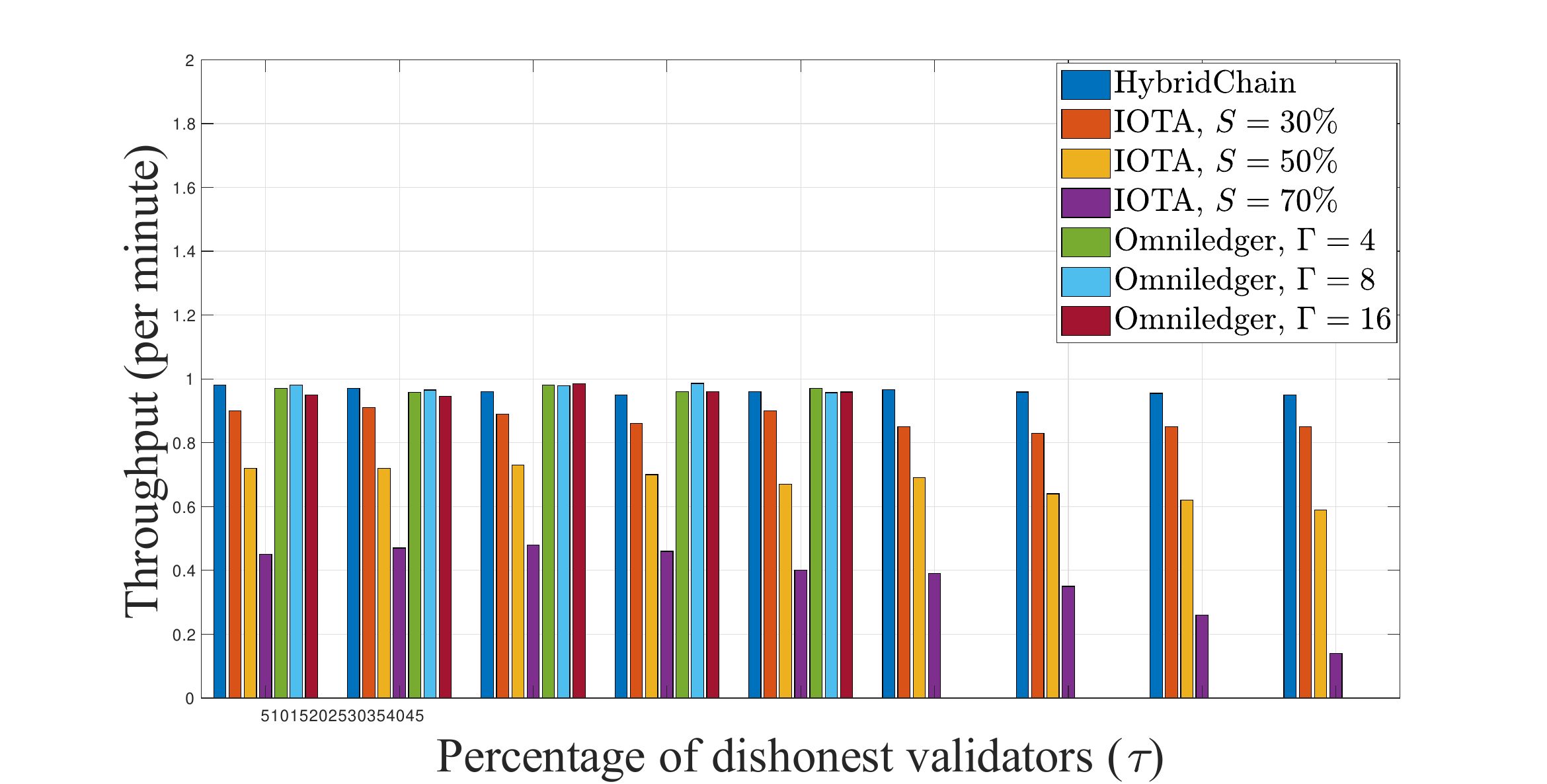}
               % \label{highcantenna333}
                }     
                }
\DeclareGraphicsExtensions.
\caption{Throughput and accuracy based on percentage of dishonest validators, with incoming transactions rate $\gamma=1200$ and $M=1000$ validators.}
\label{TTT2}
\end{figure}

The results, shown in Fig.~\ref{TTT2}, demonstrated that a decrease in incoming transactions had a more pronounced impact on IOTA, making it more vulnerable to adversary attacks. This vulnerability stems from the nature of DAG distributed ledgers, where transaction confirmation and validation rely on the structure of the DAG and other transactions. Insufficient transaction volume leads to system instability, resulting in incomplete transaction validation. Consequently, accuracy is reduced, as a greater number of valid transactions remain in the tip pool for longer periods, taking more time to be processed and becoming more susceptible to attacks. Adversaries can exploit this situation to prioritize their own transactions, hindering the processing of valid transactions and further diminishing accuracy. Specifically, by comparing the throughput and accuracy values for $\gamma=6000$ and $\gamma=1200$, it becomes evident that the trends observed for $\gamma=6000$, such as increased throughput and reduced accuracy for fixed $S$ and increased $\tau$, persist for $\gamma=1200$. However, the reduction of accuracy is more severe when $\gamma=1200$ is considered.\par
On the other hand, in HybridChain and Omniledger one transaction does not determine the validity of another transaction directly but 
, therefore changing the transaction volume does not have a significant effect on the throughput and the accuracy.\par

Hence the above results demonstrate that DAG distributed ledger systems, which do not rely on validators to process transactions, can achieve high transaction speeds, allowing network nodes to reach consensus quickly, even for large transaction volumes. However, this speed may come at the cost of consensus accuracy, particularly when facing a powerful adversary who can strategically disrupt the network.\par
\subsubsection{Scalability}
The primary objective of our last simulation is to evaluate the scalability of three schemes, a crucial element within the trilemma triangle. Scalability holds significant importance in the field of distributed ledger technologies, particularly with the proliferation of applications built on distributed platforms and the subsequent surge in user numbers and incoming transactions. In the context of distributed ledgers, scalability pertains to the system's capacity to efficiently manage an increasing volume of transactions, while sustaining high throughput and low latency. Essentially, a distributed ledger system is deemed scalable when it can process a substantial number of transactions without compromising transaction processing speed.\par
To evaluate the scalability of the three schemes, we used the settings described in Section V-B-3 while varying the rate of incoming transactions $\gamma$, from $6000$ to $3\times 10^4$ with increments of $3000$. For each $\gamma$ value, we measured the throughput of the three schemes and the maximum latency of the processed transactions. In all three schemes, a dishonest fraction of $\tau=20\%$ was incorporated. In the case of IOTA, the spamming rate $S$ was set to $30\%$, while Omniledger maintained a fixed value of $\Gamma=4$ shards. These specific parameter values were chosen based on their demonstrated superior performance in the corresponding scheme simulations conducted in Sections V-B-2 and V-B-3. The objective was to compare the scalability of the other two schemes, IOTA and Omniledger, with HybridChain unde their optimal settings.\par
\begin{figure}[!t]
\centering{
 \subfigure[t][ 
    Throughput
 ]{
                \centering
                \includegraphics[width=\columnwidth]{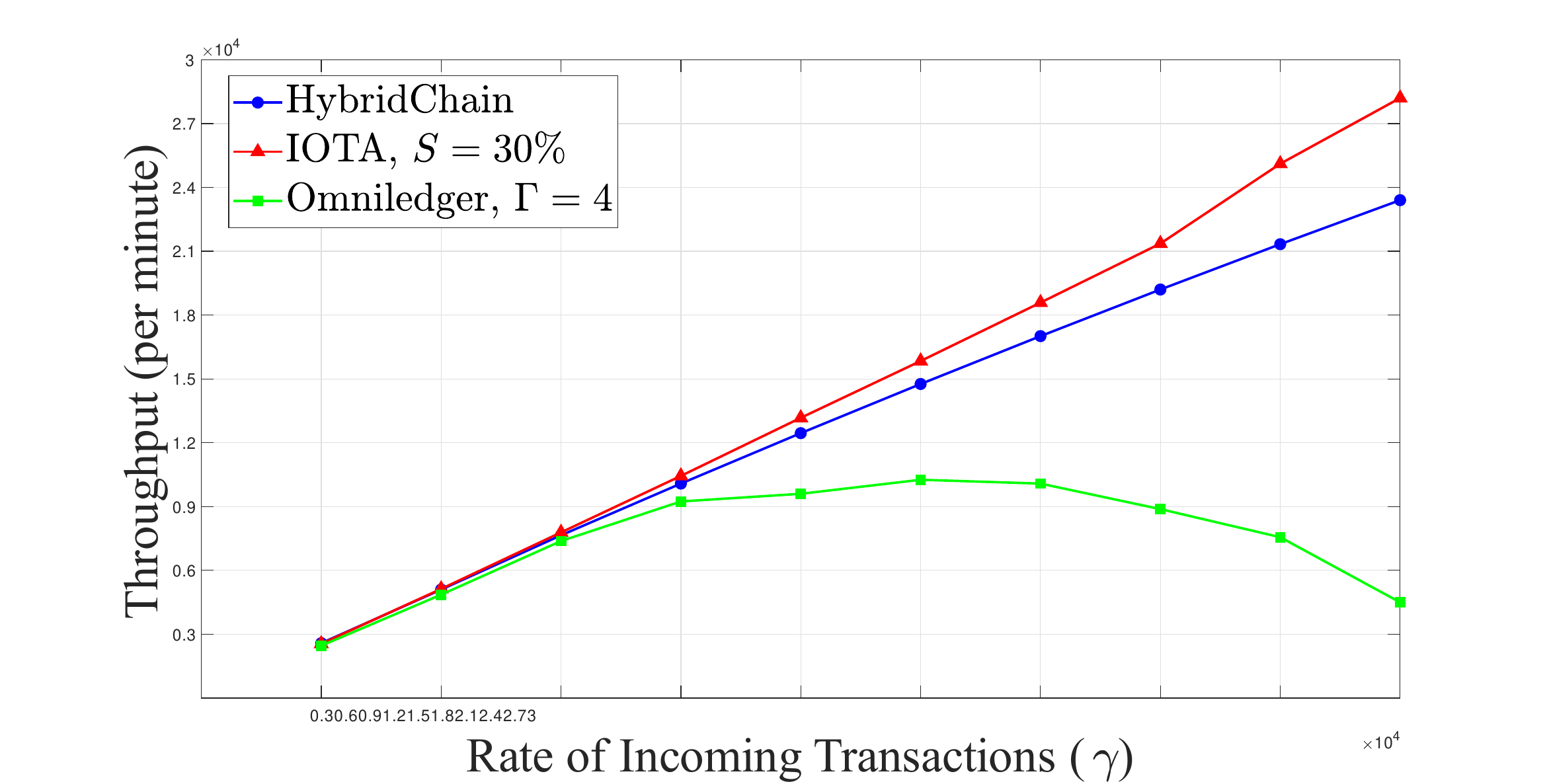}
                %\label{lowcantenna222}
                }
               \subfigure[t][ 
    Accuracy
 ]{
                \centering
                \includegraphics[width=\columnwidth]{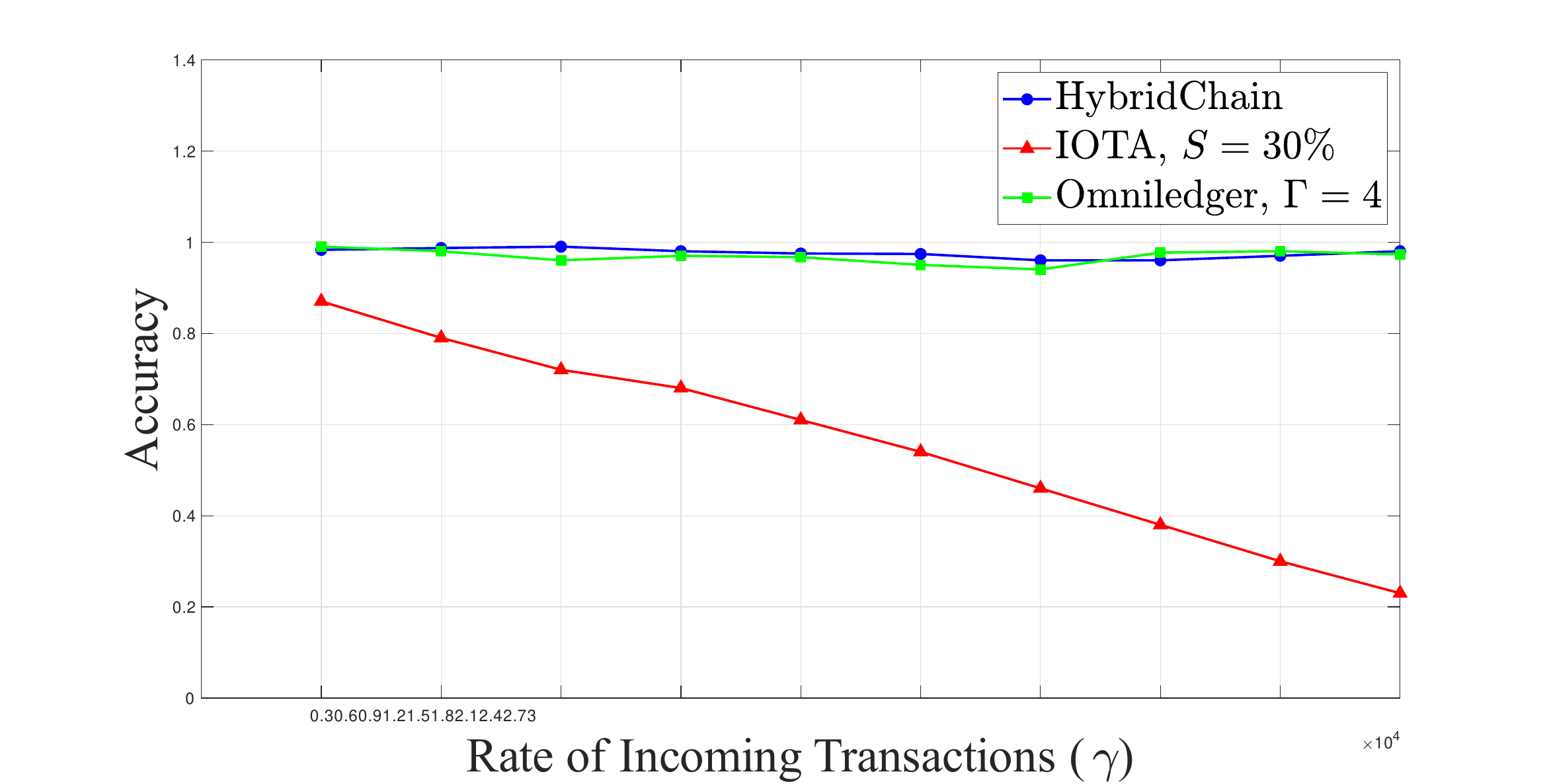}
                %\label{highcantenna222}
                }
               \subfigure[t][ 
    Maximum of Latency
 ]{
                \centering
                \includegraphics[width=\columnwidth]{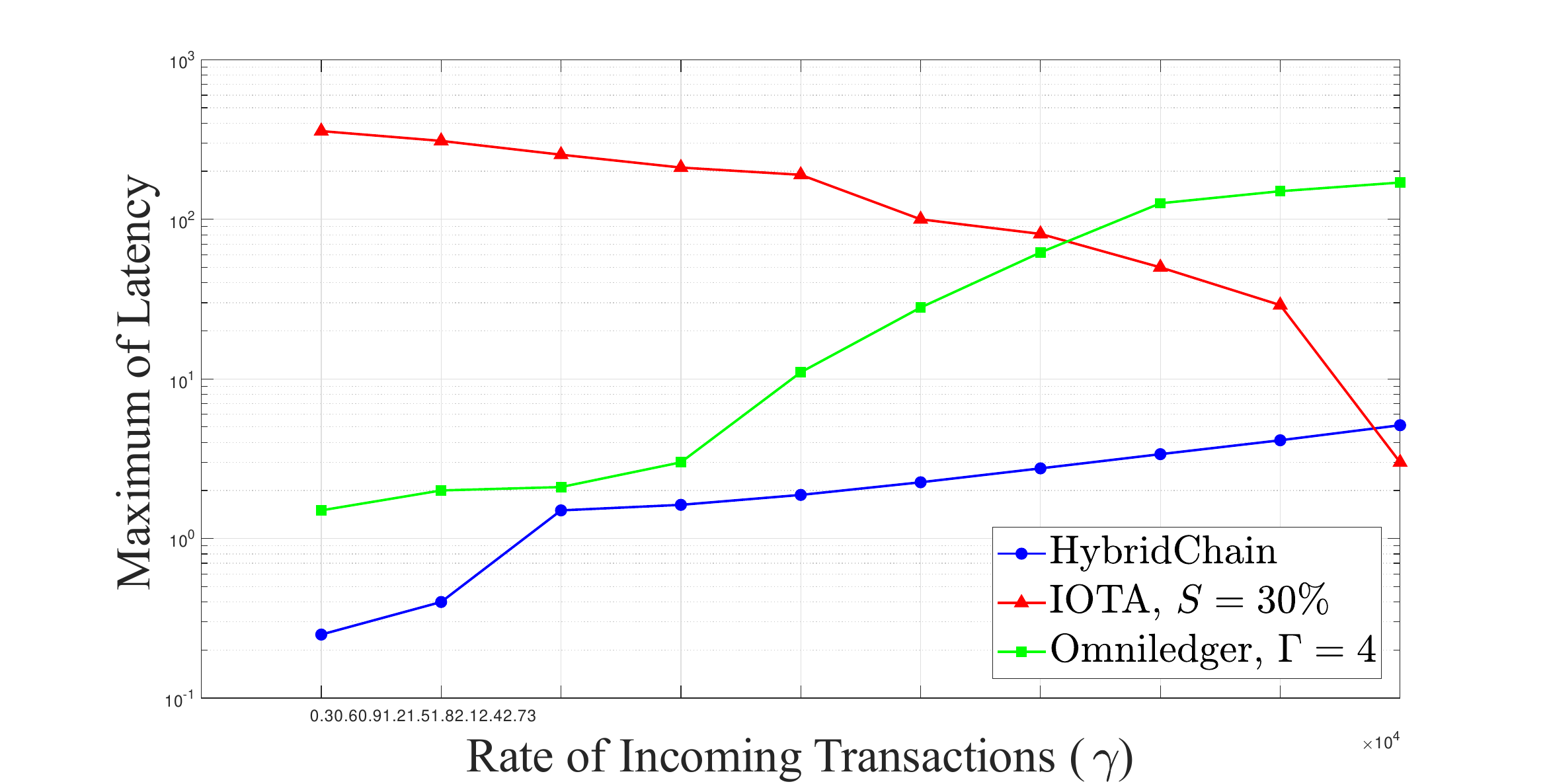}
                %\label{latency222}
                }     
                }
\DeclareGraphicsExtensions.
\caption{Scalability based on the rate of incoming transactions $(\gamma)$, with $\tau=20\%$, and $M=1000$.}
\label{SSS1}
\end{figure}

Fig. \ref{SSS1} provides a visual representation of the throughput of the three schemes, along with the maximum latency of the processed transactions. Notably, the latency trends differ among the three schemes as the rate of incoming transactions $\gamma$ increases. In IOTA, the maximum latency decreases with higher values of $\gamma$, while in the other two schemes, the maximum latency increases. This observation highlights a key characteristic of DAG schemes, where the validation of transactions relies on their relationships within the system, rather than on validators. Consequently, an increase in the rate of incoming transactions facilitates faster processing and improves throughput. However, this can potentially lead to decreased accuracy. As depicted in Fig.~\ref{SSS1}, an increase in the rate of incoming transactions $\gamma$ is accompanied by a reduction in accuracy. The absence of validators to validate transactions can indeed help reduce latency and increase throughput. However, it can also result in erroneous verifications, compromising accuracy.\par
As expected, Omniledger and HybridChain exhibit the typical behavior of non-DAG schemes, where an increase in the rate of incoming transactions $\gamma$ leads to higher maximum latency and decreased throughput due to transaction processing and communication among validators. However, the results indicate that Omniledger is more susceptible to the impact of increased $\gamma$ compared to HybridChain, as evidenced by a larger increase in maximum latency in Omniledger. HybridChain demonstrates excellent scalability, by maintaining nearly constant throughput and exhibiting only a marginal increase in maximum latency, while achieving high accuracy.\par

\section{CONCLUSIONS}
The proposed consensus scheme of HybridChain combines the strengths of blockchain and DAG, and by incorporating a decentralized learning algorithm that exploits transaction attributes, achieves fast, accurate, and secure transaction processing. HybridChain can successfully mitigate various security attacks by decentralizing storage and computations, whereas IOTA and Omniledger exhibit vulnerabilities to these attacks. Our results demonstrate that while IOTA exhibits high throughput and low latency in transaction processing, it sacrifices accuracy and becomes susceptible to orphanage attacks, particularly when the rate of incoming transactions is low. For example, when the rate of incoming transactions increases tenfold from $0.3\times 10^4$ to $3\times 10^4$ transactions per minute, IOTA's throughput increases by nearly tenfold, and the maximum latency decreases by almost $100\%$, but the accuracy declines by approximately $74\%$. In contrast, HybridChain maintains consistent accuracy, hovering around $99\%$ for all incoming transaction rates, while experiencing a maximum latency of $10.25$ seconds for a rate of $3\times 10^4$ transactions per minute. Additionally, Omniledger, as a sharded blockchain, achieves stable and acceptable accuracy by increasing the throughput through parallel processing in different shards. However, when the number of shards increases from $4$ to $16$, the maximum throughput decreases by $15\%$ due to increased latency caused by cross-shard transactions.

\end{document}